\documentclass[manuscript]{aastex} 
\usepackage{graphics}

\begin{document}
\slugcomment{\aj, in preparation (\today)}
\newcommand{\LeeZinn}{\mathscr{L}}

\title{The Globular Cluster M15. I. Identification, Discovery, \\
       and Period Determination of Variable Stars} 
\author{T.~Michael~Corwin} 
\affil{Department of Physics and Optical Science, University of North Carolina 
at Charlotte, Charlotte, NC 28223}
\email{mcorwin@uncc.edu}   

\author{J.~Borissova}
\affil{Departamento de F\'\i sica y Astronom\'\i a, Facultad de Ciencias, Universidad de Valpara\'\i so, 
Casilla 5030, Valpara\'\i so, Chile}

\email{jura.borissova@uv.cl}

\author{Peter B. Stetson}
\affil{Dominion Astrophysical Observatory, Herzberg Institute of Astrophysics, National Research Council of Canada, 5071 West Saanich Road, Victoria, BC V9E 2E7, Canada}
\email{peter.stetson@nrc-cnrc.gc.ca}

\author{M.~Catelan}
\affil{Pontificia Universidad Cat\'olica de Chile, Departamento de
       Astronom\'\i a y Astrof\'\i sica, \\ Av. Vicu\~{n}a Mackenna 4860,
      782-0436 Macul, Santiago, Chile}
\email{mcatelan@astro.puc.cl}

\author{Horace A. Smith}
\affil{Dept.\ of Physics and Astronomy, Michigan State University,
       East Lansing, MI 48824}
\email{smith@pa.msu.edu}

\author{R. Kurtev}
\affil{Departamento de F\'\i sica y Meterolog\'\i a, Facultad de Ciencias, 
Universidad de Valpara\'\i so, Casilla 5030, Valpara\'\i so, Chile}
\email{rkurtev@uv.cl}

\and
\author{Andrew W. Stephens}
\affil{Gemini Observatory, 670 N. A'ohoku Place, Hilo, HI 96720}
\email{stephens@gemini.edu}

\begin{abstract}
We present new $BVI$ CCD photometry for variables in the globular cluster M15. Our photometry was obtained using both the image subtraction package ISIS and DAOPHOT/ALLFRAME. The data were 
acquired in 2001 on two observing runs on 11 observing nights using the 2-m telescope of the Bulgarian National Astronomical
Observatory ``Rozhen'' with a Photometrics CCD camera. For 40 
previously known variables, we present a period for the first time, and improved periods 
were obtained for many previously known variables.  Fourteen new variables are reported. We present updated Bailey diagrams for the cluster and discuss its Oosterhoff classification. 
Although many of M15's RRab pulsators fall at an intermediate locus between Oosterhoff types 
I and II in the Bailey diagram, we argue that M15 is indeed a bona-fide Oosterhoff type II 
globular cluster.

\end{abstract}


\section{Introduction}
 
The metal-poor globular cluster M15~=~NGC~7078 (${\rm [Fe/H]} = -2.26$; 
Harris 1996) was one of the 
three original ``type II'' (OoII) clusters identified in Oosterhoff's (1939) seminal paper. 
It is a post-core-collapse cluster with an extraordinarily dense center (Sosin \& 
King 1997). Observations of its RR Lyrae variables have long been seen as important 
to clarifying the origin and nature of the Oosterhoff phenomenon. M15 is known to 
contain over 180 intrinsic variables, predominately RR Lyraes (Clement et al. 2001
and references therein). 

RR Lyrae variables provide crucial information for estimating globular cluster ages 
and distances, as summarized by Smith (1995). They are easily identified by their 
distinctive light curves and are bright enough to be observed to considerable 
distances. Their absolute magnitudes appear to be quite restricted. The range of 
RR Lyrae luminosities is discussed, among others, in Carney (2001), Harris (2001), and Catelan (2007). Stellar 
variability studies of globular clusters are also of fundamental importance to 
understanding both stellar and cluster evolution, as well as to constrain the formation history of the Galaxy and its nearby satellites (Catelan 2007). 

The variables in M15 were previously studied by Sandage, Katem, \& Sandage (1981), Bingham et al. (1984), Silbermann \& Smith (1995), Butler et al. (1998), \'{O}~Tuairisg et al. (2003, hereafter 
OT03), Zheleznyak \& Kravtsov (2003, hereafter ZK03), and Arellano Ferro, Garc\'{i}a Lugo, \& Rosenzweig (2006)  among others. In the 
present paper, we present the results of a new variability analysis of the 
cluster, carried out using both image-subtraction techniques and the standard DAOPHOT/ALLFRAME analysis. We provide periods for 39 previously known variables whose periods were
unknown, improved periods for 30 previously known variables, and periods for 14 new variables. In \S2, we discuss 
our observational material and reduction procedures. In \S3, we present our 
periods and derived light curves in either absolute $B$ magnitudes or relative $B$ fluxes. \S4 discusses standard magnitudes and light curve amplitudes, as well as the cluster's Oosterhoff classification, while \S5 contains notes on 
individual stars. We close in \S6 by summarizing our main results. 

\section{Observations and Data Reduction}
\subsection{Observations}

Time-series photometry was obtained during the interval 2001 July 13 to 2001 
July 19 with the 2-m Ritchey-Chr\'etien telescope of the Bulgarian National Astronomical 
Observatory ``Rozhen'' with a Photometrics CCD camera. This represents almost all of 
our data. During the interval 2001 September 18 to 2001 September 25, using the 
same telescope and instrument, a few additional images were obtained. The scale 
at the Cassegrain-focus CCD was 0.33 arcsec per pixel and the observing area was 
centered on the cluster center. In total 230 $B$, 201 $V$, and 163 $I$ frames were obtained, 
with seeing between 0.8 and 1.2~arcsec. In order to resolve better the cluster 
center, we observed M15 in two different CCD modes, unbinned $1024 \times 1024$ CCD 
format in the nights with the best seeing and in binned $512 \times 512$ CCD format on 
the other nights.

\subsection{Data Reduction}

The raw data frames were processed following standard procedures to remove the 
bias, trim the pictures, and divide by mean sky flats obtained using color-balanced 
filters. No attempt was made to recover bad pixels or columns. We employed the 
image subtraction package ISIS V2.1 (Alard \& Lupton 1998; Alard 2000) on these images. The 
resulting differential flux data produced light curves for a large number of 
variables. These stars were initially matched with those in the Clement et al. 
(2001) online catalog, and later as OT03 
and ZK03 came out, with their newly discovered variables. In this analysis 13 new variables were discovered. A later DAOPHOT/ALLFRAME analysis (Stetson 1994) on the same data resulted in better astrometry. An additional variable was detected in this analysis.

\def\hr{${}^{\rm h}\,$}
\def\mn{${}^{\rm m}\,$}
\def\Sc{${}^{\rm s}$\llap{.}}
\def\Deg{${}^\circ$\llap{.}}
\def\Sec{${}^{\prime\prime}$\llap{.}}
\def\deg{${}^\circ\,$}
\def\min{${}^{\prime}\,$}

For the DAOPHOT/ALLFRAME analysis, the following procedure was used 
to determine the RA and Dec of the variable stars. Absolute equatorial 
coordinates (equinox J2000.0) for 11,738 stars within a
1\Deg0\,$\times$\,1\Deg25 (E-W\,$\times$\,N-S) field centered on position
21\hr29\mn34\Sc15 +11\deg59\min15\Sec0 were selected from the USNO-A2.0 Guide
Star database [Monet et al. 1998, USNO-A2.0 (Flagstaff: US Naval Obs.), CD-ROM]
maintained by the Canadian Astronomy Data Centre. The equatorial coordinates
were transformed to $(\xi,\eta)$ rectangular coordinates measured in arcseconds
relative to the reference position just given. At the same time, square
1\deg\,$\times$\,1\deg images were extracted from the Digitized Sky Survey I
``O'' data and the DSS-II $B$, $R$, and $I$ data, likewise through the
services of the Canadian Astronomy Data Centre. Positions and magnitude
indices for stars detected within these digitized Schmidt plates were obtained
with a modernized version of the codes described in Stetson (1979a,
1979b). The program DAOMASTER was used to transform the
photographic positions together with those from our CCD reductions to the
$(\xi,\eta)$ coordinate system defined by the USNO Guide Star catalog. From
there, they could be transformed to the J2000.0 equatorial system of the
USNO-A2.0 catalog. We anticipate that our absolute positions for {\it
uncrowded\/}, well-exposed stars are accurately on the mean USNO-A2.0 system to 
well under 0\Sec1, but
positions of very faint stars or blended stars are likely to be more uncertain
than this.

\begin{figure*}[t]
  \figurenum{1}
  \epsscale{0.99}
\plotone{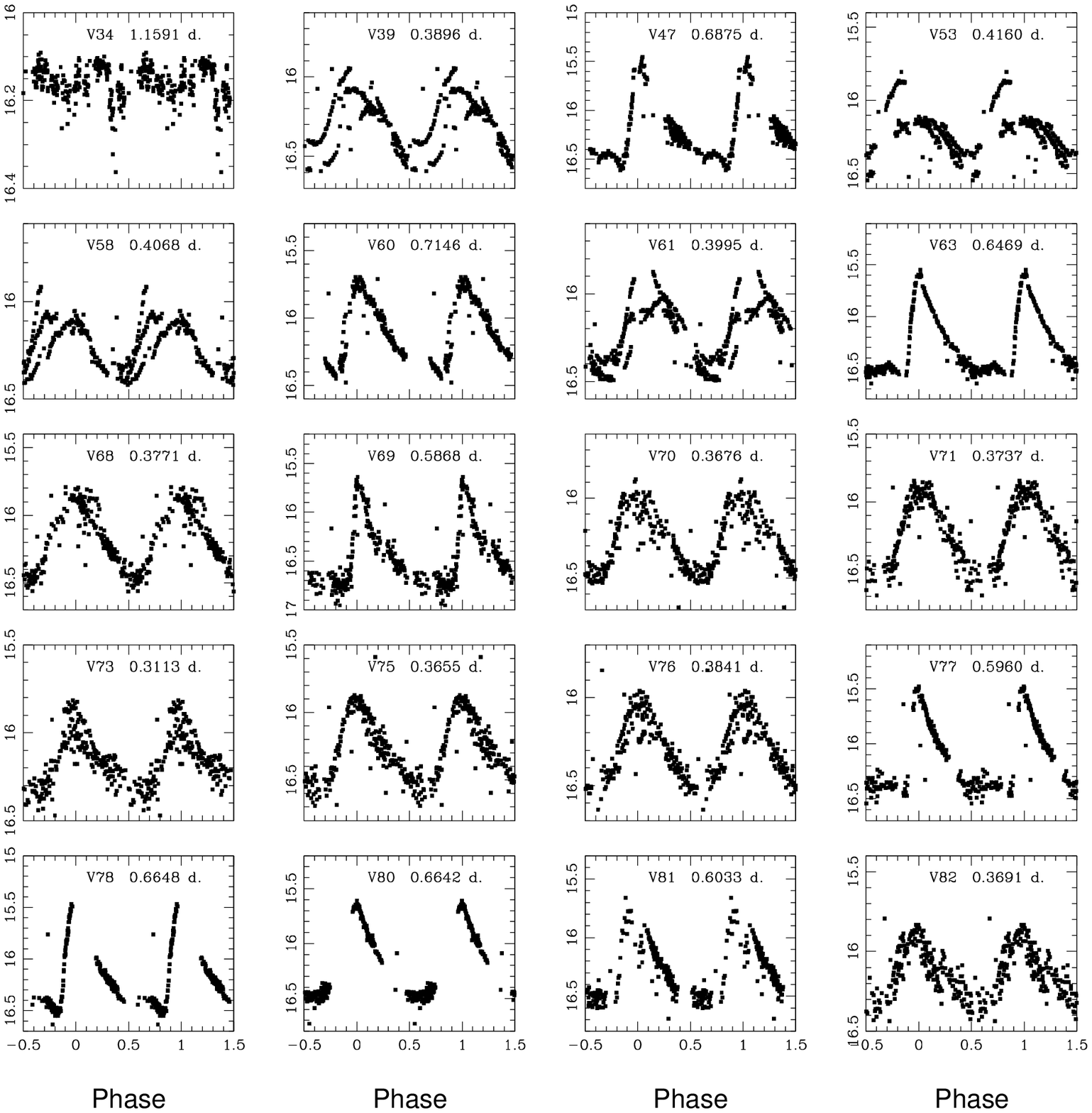}
  \caption{$B$ light curves for selected previously identified variable stars in M15. ISIS data are plotted in arbitrary units and values are not listed on the y axes. DAOPHOT/ALLFRAME data are plotted in standard magnitudes.
      }
      \label{Fig01a}
\end{figure*}

\begin{figure*}[t]
  \figurenum{1\,({\em continued})}
  \epsscale{0.99}
\plotone{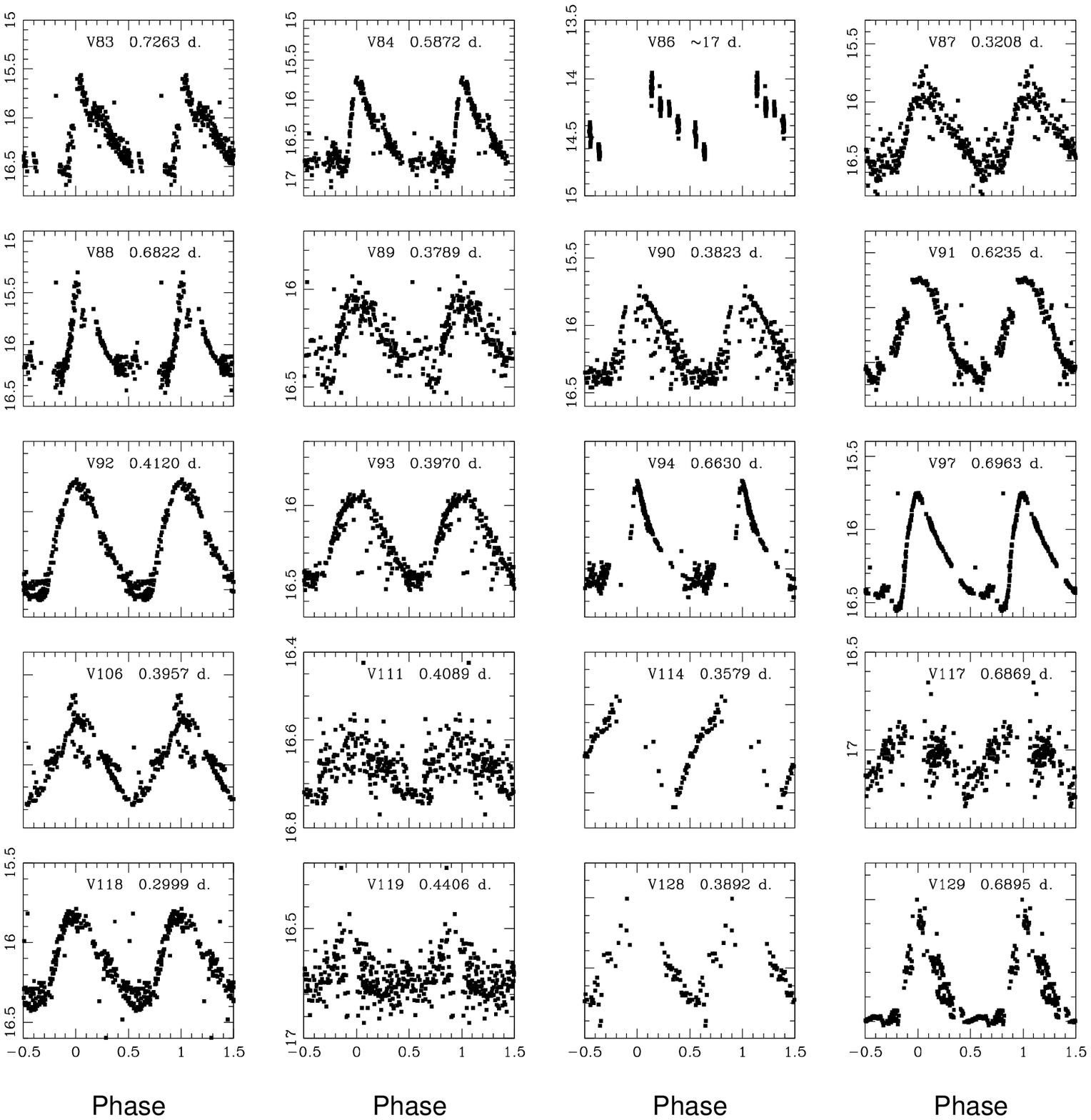}
  \caption{}
      \label{Fig01b}
\end{figure*}

\begin{figure*}[t]
  \figurenum{1\,({\em continued})}
  \epsscale{0.99}
\plotone{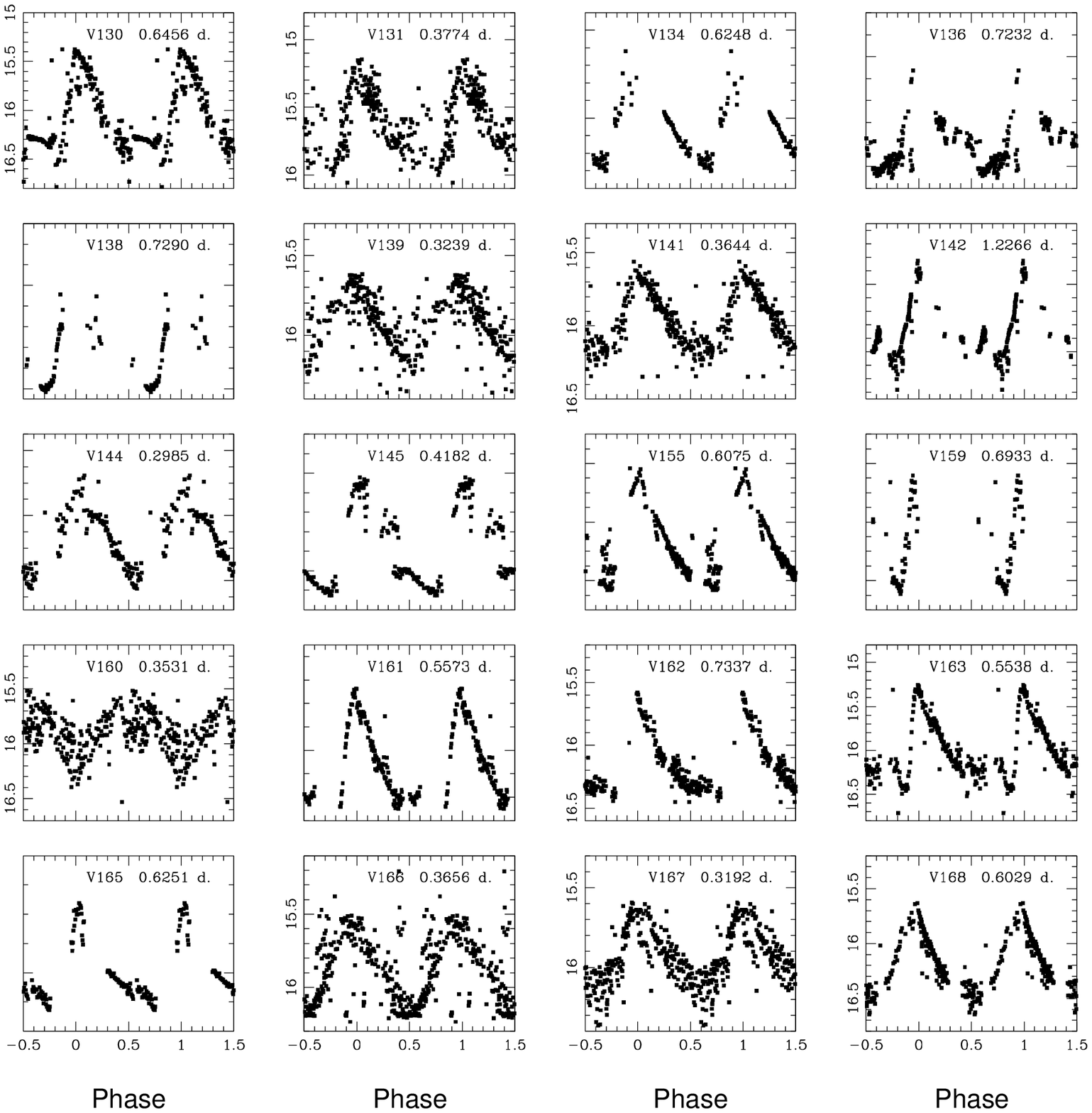}
  \caption{}
      \label{Fig01c}
\end{figure*}

\begin{figure*}[t]
  \figurenum{1\,({\em continued})}
  \epsscale{0.99}
\plotone{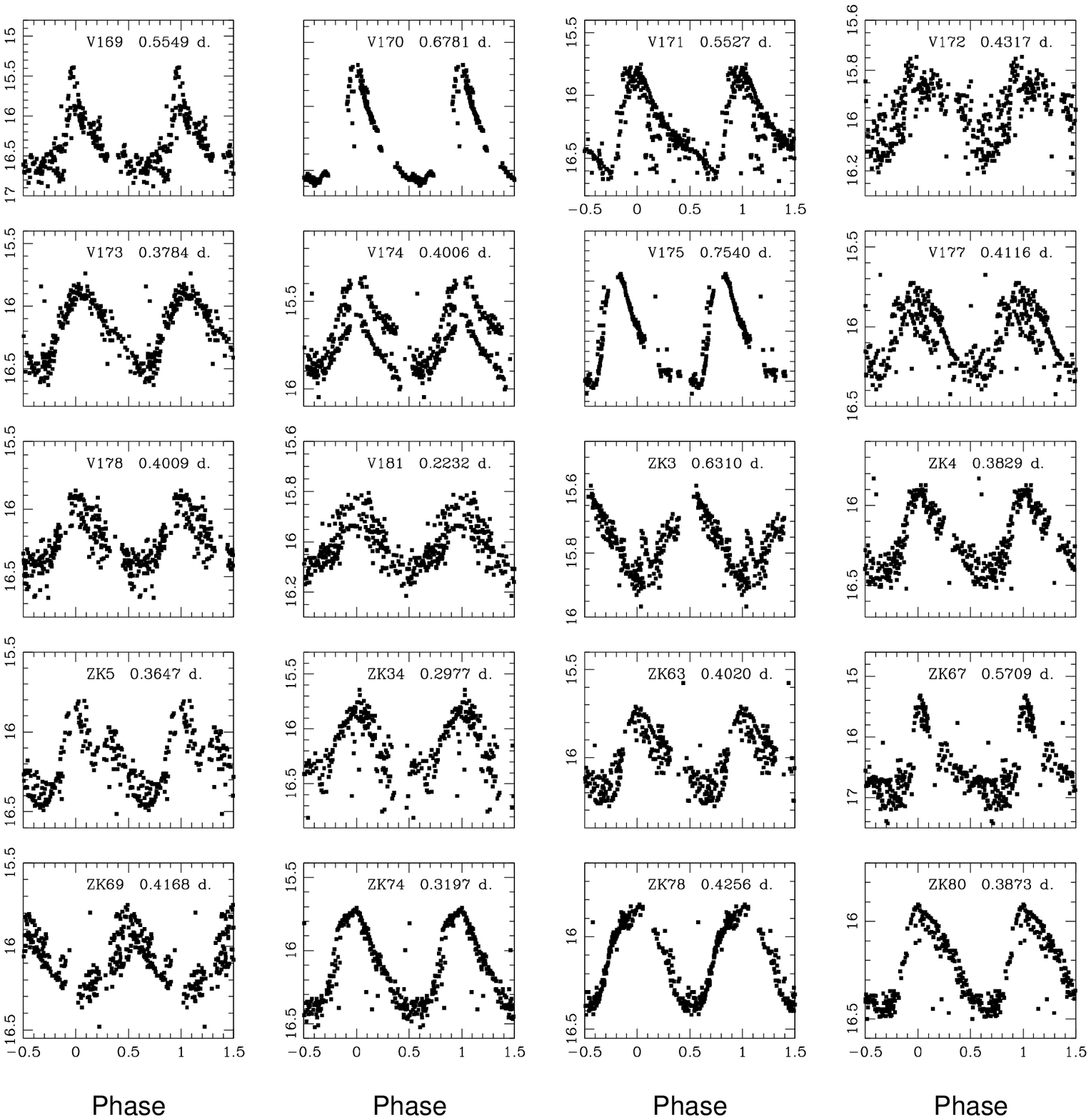}
  \caption{}
      \label{Fig01d}
\end{figure*}

\begin{figure*}[t]
  \figurenum{2}
  \epsscale{0.85}
\plotone{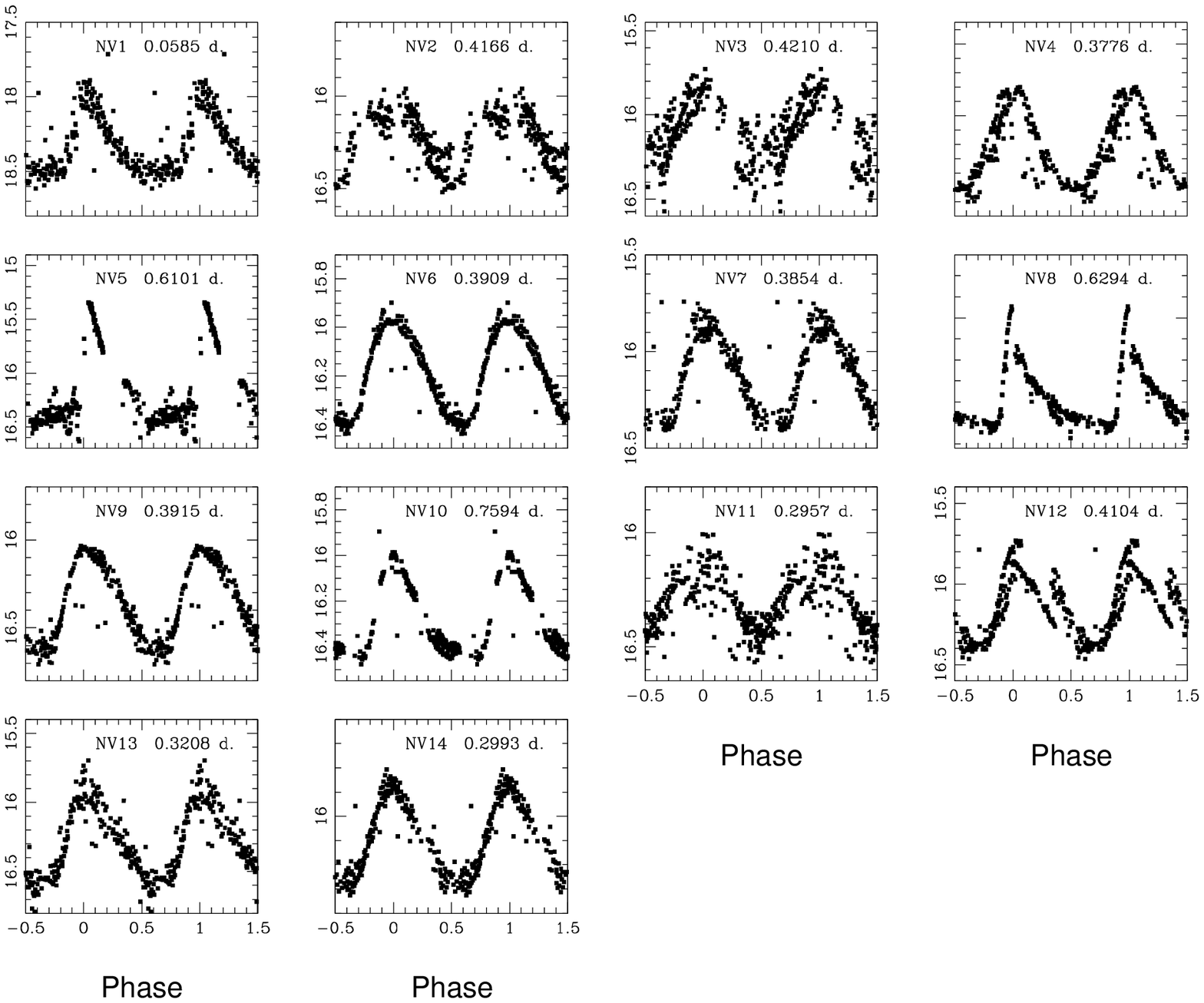}
  \caption{$B$ light curves for the newly discovered variable stars in M15. ISIS data are plotted in arbitrary units while DAOPHOT/ALLFRAME data are plotted in standard magnitudes.
      }
      \label{Fig02}
\end{figure*}

\begin{figure*}[t]
  \figurenum{3}
  \epsscale{0.85}
\plotone{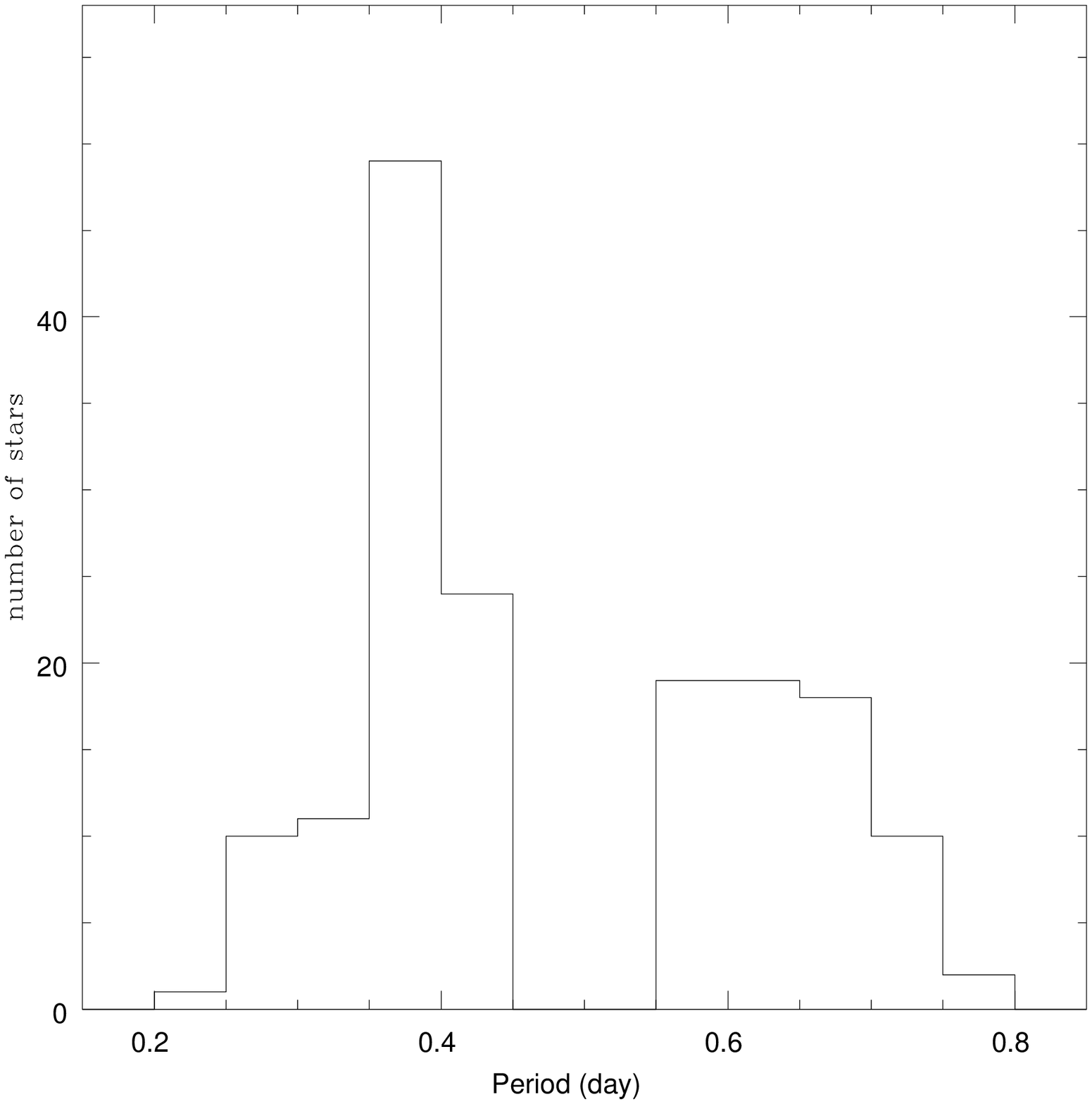}
  \caption{Histogram of the periods of M15 variables. 
      }
      \label{Fig03}
\end{figure*}

\begin{figure*}[t]
  \figurenum{4}
  \epsscale{0.85}
\plotone{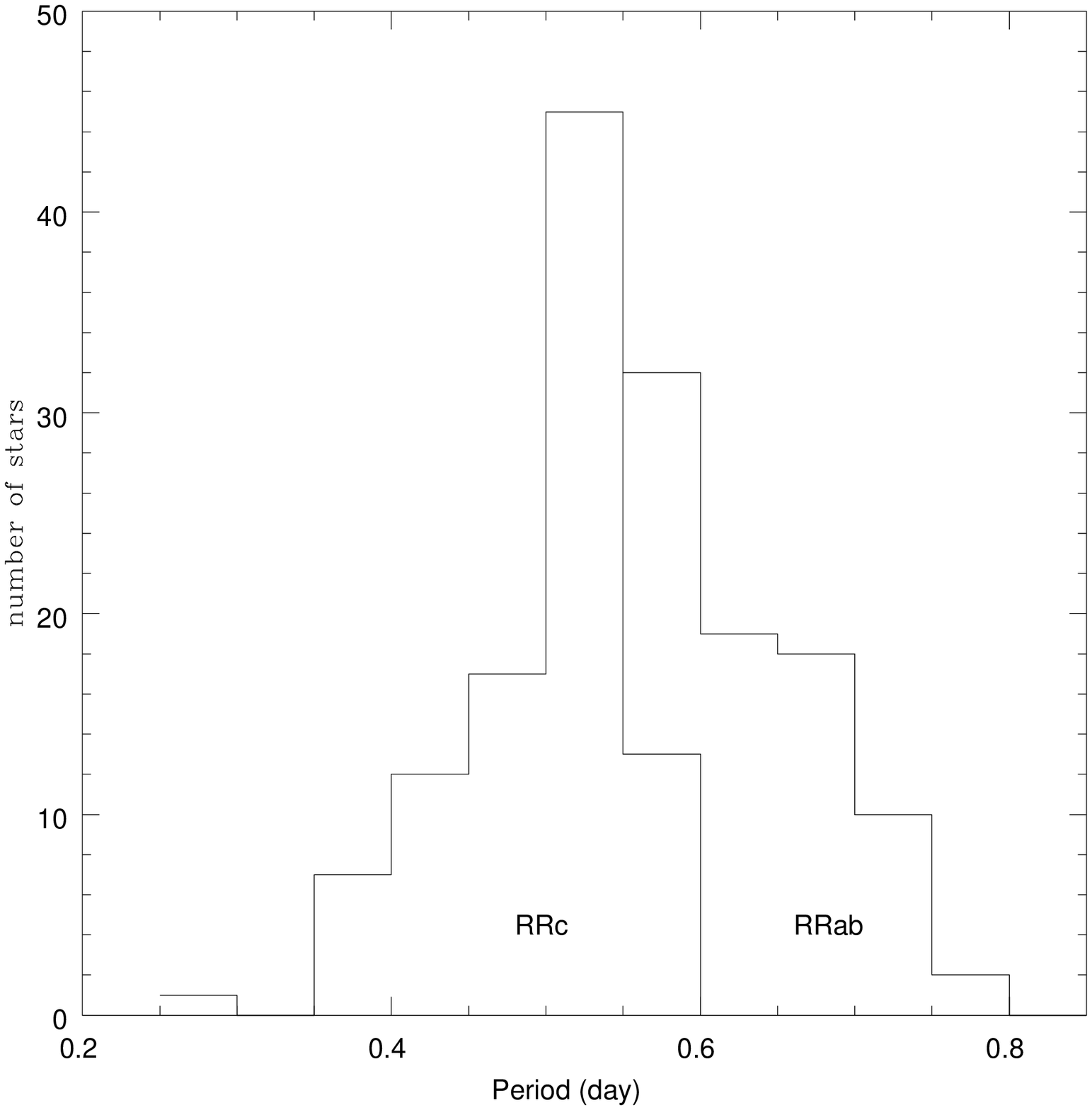}
  \caption{Histogram of the fundamentalized periods of M15 variables. 
      }
      \label{Fig04}
\end{figure*}

\begin{figure*}[t]
  \figurenum{5}
  \epsscale{0.85}
\plotone{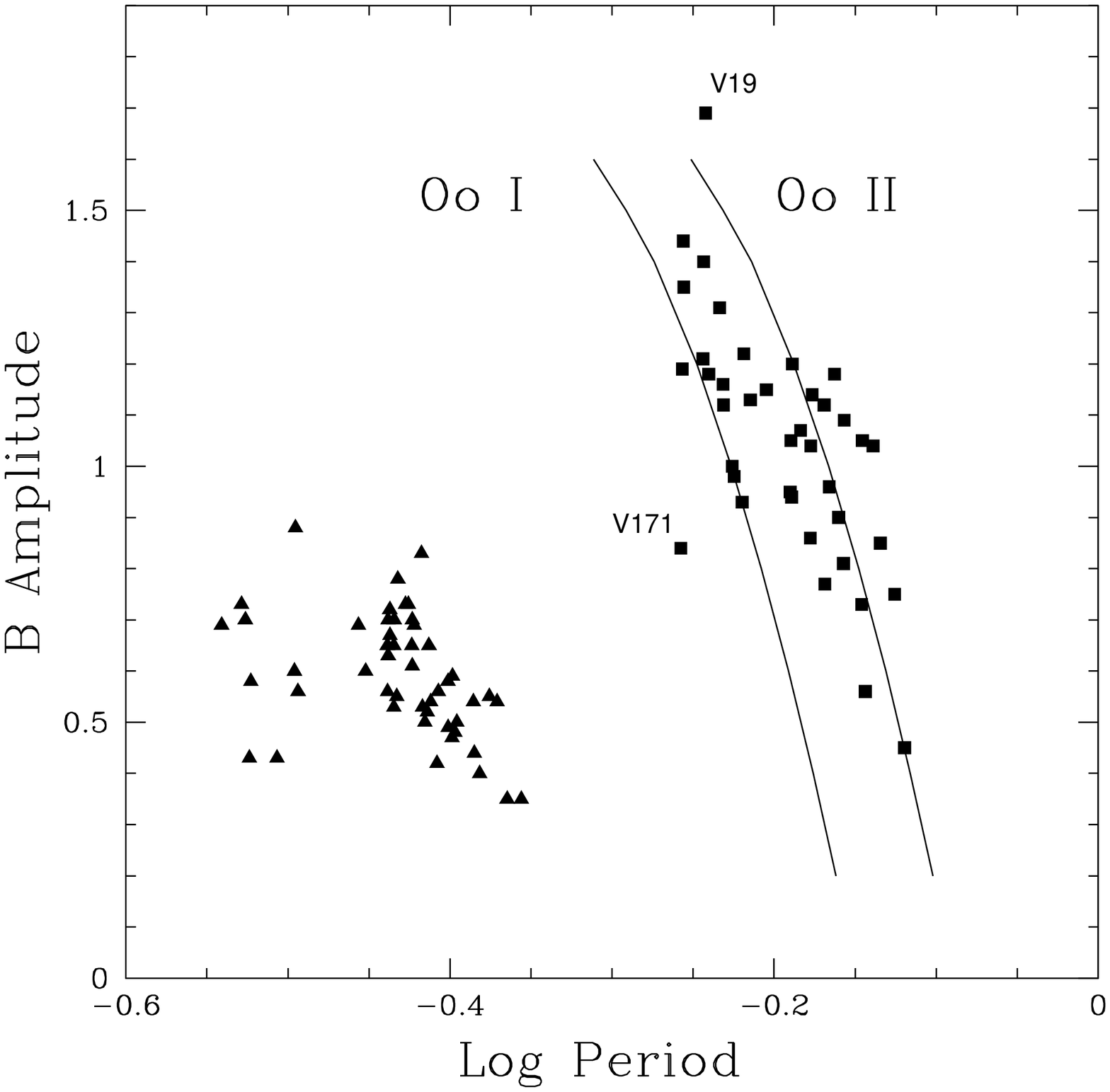}
  \caption{$B$ amplitude vs log period (days). Filled triangles for RRc and filled squares for RRab. 
      }
      \label{Fig05}
\end{figure*}

\begin{figure*}[t]
  \figurenum{6}
  \epsscale{0.85}
\plotone{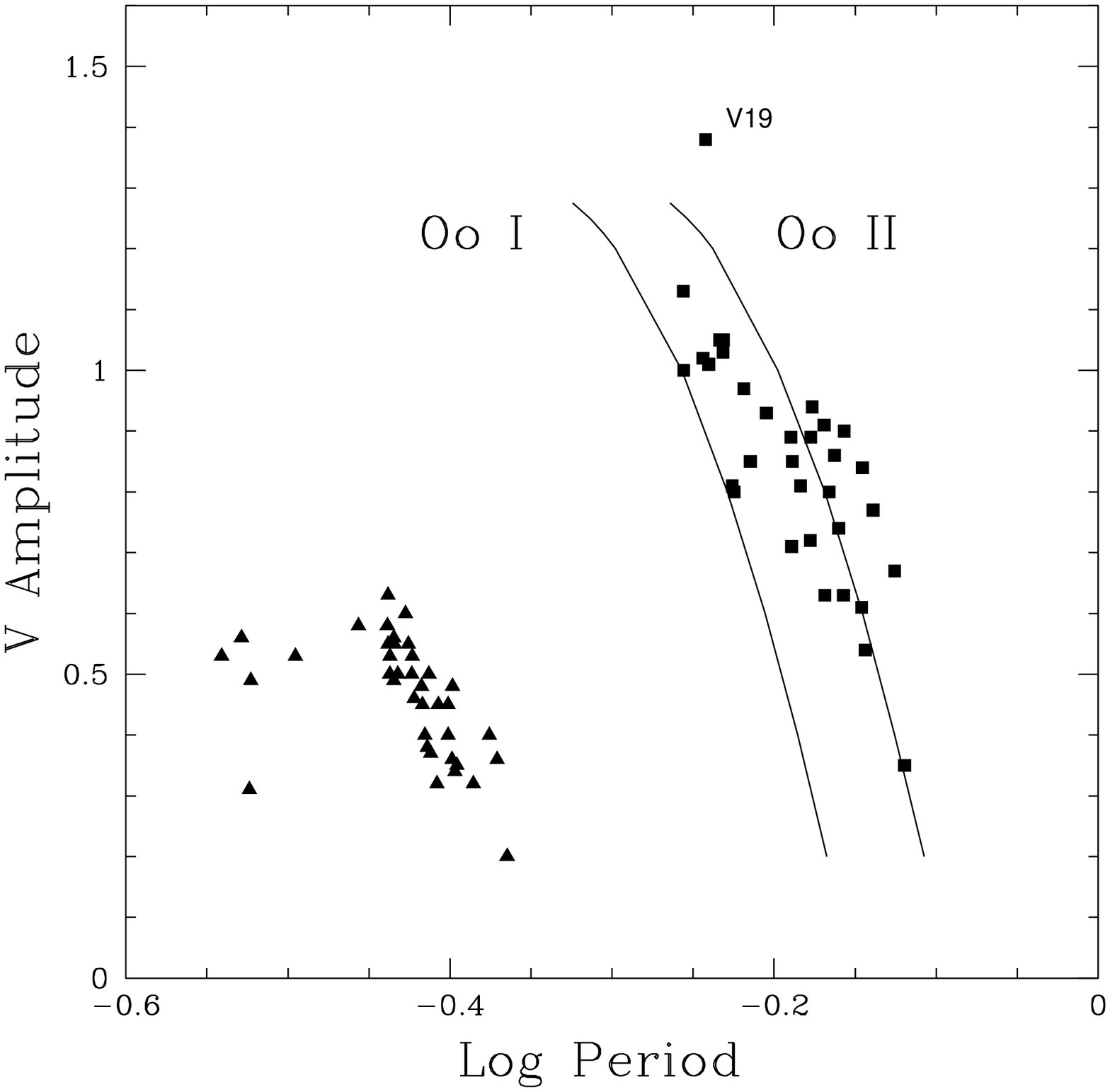}
  \caption{$V$ amplitudes vs. log period (symbols as in Figure 5). 
      }
      \label{Fig06}
\end{figure*}

\begin{figure*}[t]
  \figurenum{7}
  \epsscale{0.85}
\plotone{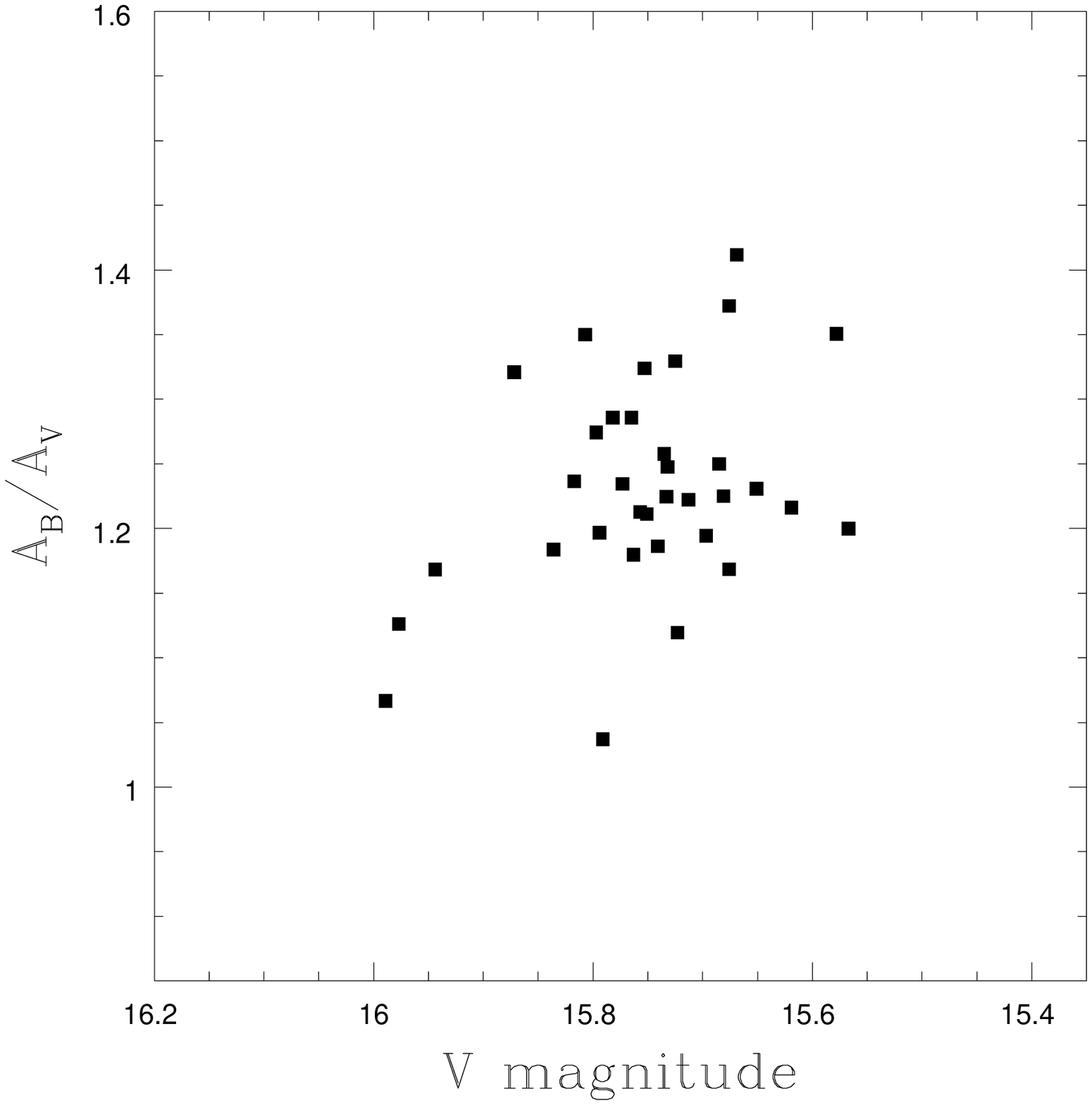}
  \caption{$A_B/A_V$ vs. $V$ magnitude. 
      }
      \label{Fig07}
\end{figure*}

\section{Periods and Light Curves}

Table~1 gives the location of all the M15 variables found in this study in RA and Dec. The positions come from the research program described in Stetson (2000). Table~1 includes previously published periods when they exist and our newly determined periods. The variable type is also given in Table~1. 

Most of our data cover 7 days, spanning about 21 cycles for the shorter-period variables and about 10 cycles for the longer-period variables. There are a few additional data extending the time sequence to about 70 days. Because of the relatively short time interval of our data, periods are quoted to only 4 significant figures. Assuming we have correctly identified the variable, in several cases our period differs significantly from those previously published. In each case, the published period will not phase our data. 

Table~2 gives the location of selected variables, primarily recent discoveries, generally toward the center of the cluster. In Table 2 we have converted RA and Dec coordinates to arcsec from the center of the 
cluster, as in the Sawyer-Hogg (1973, hereafter S-H) and the Clement et al. (2001) catalogs. The procedure we used for obtaining those values is as follows. Our coordinates 
in RA and Dec (as shown in Table 1) were converted to the S-H system by matching 
many of our variables with known variables from the catalog. The coordinates in 
OT03 are also given in RA and Dec. Similarly, these were converted to the S-H 
system by matching the previous known variables given at the top of their Table~4 
with their corresponding values in the Clement catalog. The coordinates in 
ZK03 are given in the S-H system. ZK03 also provide corrected S-H 
coordinates for variables 128 through 155. Table 2 is intended to indicate the reliability of our identifications of previously known variables.

Periods were determined using the period-search program ``kiwi,'' which is
based on the Lafler \& Kinman (1965) algorithm. In other words, ``kiwi''
searches for periodicity by seeking to minimize the total length of the line
segments that join adjacent observations in phase space, i.e., to maximize the
smoothness of the light curve. (The ``kiwi'' program was kindly provided to us by Dr. Betty Blanco.) 

Selected light curves based on our periods given in Table~1 are shown in Figures 1 and 2. Figure~1 includes some previously reported variables, but light curves for well established variables are not plotted. Figure~2 includes the variables 
reported for the first time in this paper. Most light curves are plotted in standard $B$ magnitudes, while for some variables ISIS differential flux curves were used. In general, the light curves for a given star are smoother for the ISIS data than for the DAOPHOT/ALLFRAME data. The reason is that the ISIS reduction procedure consists of several steps. Initially, all the frames are transformed to a common coordinate grid. Next, a composite reference image is created by stacking several best seeing frames. For each frame, the composite reference image is convolved with a kernel to match its PSF and then subtracted. On the subtracted images, the constant stars cancel out, and only the signal from variable stars remains. A median image is constructed of all the subtracted images, and the variable stars are identified. Finally, profile photometry is extracted from the subtracted images. Thus, using ISIS we have obtained the relative fluxes for every variable star without contamination of the neighbor constant star companions and with much smoother, if any, sky background. 
Because of the standard magnitudes available in the DAOPHOT/ALLFRAME data, we chose to use these data for the figure whenever possible. ISIS data are used only for variables for which there were no DAOPHOT/ALLFRAME data or for which those data were extremely noisy. The internal precision of the photometry covers a wide range. For many stars 
the scatter about the mean loci is relatively small. For others it is barely 
sufficient to show that the star is in fact variable. 

As already stated (\S2.1), our data consist of images taken in two different formats. Nights with heliocentric 
Julian dates from 2452104 through 2452109 utilized a $512 \times 512$ CCD while night 
2452110 
and the later nights utilized a $1024 \times 1024$ CCD. Our initial image subtraction 
analysis utilized two different sets of reference frames, one $BV$ set for each of the 
image formats. The fluxes are calculated with respect to these two different reference frames and as a result we have two different flux systems. In principle, it should be possible to transform the relative fluxes to the instrumental magnitudes of the reference frames and thus to the standard system. The variables in the very crowded parts of the cluster, however, have no measured instrumental magnitudes. We also noticed some non-linearity in the transformation (see for example, Borissova et al. 2001). We then chose to rebin the original $512 \times 512$ images
 to $1024 \times 1024$ 
images. This allowed us to use a single $B$ and a single $V$ reference frame for the 
entire range of data. It is this analysis that is reported in this paper.

The comparison between original and rebinned data shows that the light curves of 
rebinned data are somewhat noisier. Since the rebinning conserves flux, there are two possible 
sources of noise. One is the difference between the real PSF and the very circular 
PSF which results from linearly interpolating the undersampled 512 images. The second 
is a much lower signal to noise ratio for the sky. However, having one single reference 
frame per filter allowed us to obtain a single differential flux light curve per filter for each variable star. 

\section{Standard Magnitudes and Light Curve Amplitudes}

Using the light curves produced by the DAOPHOT/ALLFRAME analysis, the magnitude-averaged and intensity-averaged mean magnitudes of the variables were determined. These are shown in Table 3. Also shown in Table 3 are the BVI amplitudes. These were determined from measurements made on hardcopies of the light curves. As a check of the accuracy of this approach, we fitted some of the light curves with a high order polynomial function and calculated the amplitude of the variable. The error in amplitude was determined by adding in quadrature the standard deviation of the light curve fit at maximum and minimum light. These calculations indicate that the errors of the amplitudes are within 0.01 to 0.03 mags. For the light curves that were very noisy, including many of the suspected double-mode pulsators, no amplitudes are given.

Figure~3 is a histogram of the M15 RR Lyrae periods with the RRab (or RR0) and RRc (RR1) variables 
clearly separated.  For the purposes of this figure, the suspected RRd (RR01) variables have been grouped with the RRc variables. The RRc variables (including possible RRd variables) all have periods less than 0.45~d while the 
RRab variables all have periods greater than 0.55~d. The average period for the 
68 stars that are very likely RRab variables is 0.644~d and for the 95 stars that are very likely RRc (or RRd) variables is 0.370~d. Figure~4 is similar to Figure~3 except 
that the periods of the RRc (and possible RRd) variables have been fundamentalized by adding 0.128 to 
their log period. As can clearly be seen from this plot, the fundamentalized period distribution presents a clear peak at a value $P_{\rm f} \simeq 0.53$~d. A secondary ``hump'' is also hinted at, located at $P_{\rm f} \simeq 0.67$~d. Both these features were previously noted and discussed in Catelan (2004), who points out that neither of these features is predicted by canonical stellar evolution/pulsation theory. In particular, such peaked distributions appear especially difficult to account for in terms of the evolutionary paradigm for the origin of OoII clusters, according to which RR Lyrae variables in the latter are evolved away from a position on the blue zero-age HB. Indeed, the average period of the RRab variables (0.644~d), the average periods of the RRc variables (0.370~d), the number fraction of c-types (1.40), and the minimum period of the RRab-type variables (0.55380~d) all consistently indicate an OoII type for the cluster. As recently shown by Catelan et al. (2007), $\langle P_{\rm ab}\rangle$ and $P_{\rm ab,min}$ are particularly reliable indicators of Oosterhoff type.

It is somewhat surprising, therefore, that M15's RRab stars do not appear to follow the same trend as do RRab stars in other globular clusters, as far as the Bailey diagram is concerned. This is shown in Figure~5, which provides a plot of $B$ amplitude as a function of log period. As has long been known, for the RRab variables, the amplitude decreases as the period increases, while for the RRc variables, the amplitude first increases and then decreases with increasing period (Sandage 1981a). Also shown in the Figure are typical lines for the RRab variables in Oo II and Oo I clusters. These were taken from Cacciari, Corwin, \& Carney (2005). Figure~6 is a plot of $V$ amplitudes as a function of log period. 
As can clearly be seen, there are many stars in the cluster occupying an intermediate position in the Bailey diagram between OoI and OoII, and even some that better match the OoI locus. This has been seen before for M15, for instance in Figure 4 of Cacciari et al. 2005. Also, as shown by Pritzl et al. (2002), there is indeed no a priori reason why all stars in all Oo II globular clusters should fall precisely on the same line (see their Fig.~16), so that the apparent ``Oosterhoff-intermediate'' nature of M15 in the Bailey diagrams of Figures~5 and 6 may in fact reveal more a deficiency of the latter in correctly identifying Oosterhoff status than an actual intermediate status for the (prototypical!) type II globular cluster M15---a cluster which, as already stated, completely lacks RRab variables with periods shorter than 0.55~d, which happens to be the typical {\em average} RRab period in OoI globular clusters (e.g., Smith 1995; Clement et al. 2001; Catelan 2007).

A factor that would affect the location of RR Lyrae stars in the Bailey diagram is the presence of blended stars in the data. Blended stars have lower amplitudes for a given period. In these cases, a plot with the correct (unblended) amplitude would move the variable up in the diagram, in the direction of the typical line for OoII clusters. A blend with a star of different color (which in general one would expect to be the case) would also affect the ratio of the $B$ and $V$ amplitudes. To look for this effect, we plotted the ratio of $A_B/A_V$ against $V$ magnitude (see Figure 7). However, in the figure there appears to be very little difference in the amplitude ratios of the brighter and the fainter stars. Higher resolution images of M15 taken with the $Hubble Space Telescope$ (to be discussed in Paper II) have the potential to reveal near neighbor stars that might go undetected in observations made from the ground. However, inspection of HST images of M15 did not reveal a significant number of likely blends. At this time, we have no clear explanation for the period shift spread in the M15 RRab variables.

The location of an RRab variable in the period-amplitude plane of Figure~5 is a 
measure of its average luminosity, which is determined by its surface temperature 
and radius (the Stefan-Boltzmann equation). Sandage (1981a, 1981b), Jones et al. 
(1992), Catelan (1998), and Sandquist (2000) have shown that the $B$ amplitude of 
an RRab variable is related to its effective temperature; the larger the amplitude, 
for a given metallicity, the higher the temperature (but see De Santis 2001). The 
pulsation equation relates the period of a variable star to its average density; 
the period is inversely proportional to the square root of the average density. 
Assuming that the masses of the variables are distributed in a narrow range, the 
period becomes a measure of the average size of the star; the longer the period, 
the larger the star. Thus, for a given amplitude (temperature), the longer period 
(larger) variables should be more luminous.
Assuming all of the variables are cluster members and thus at about the
 same distance, 
the position of a variable in the period-amplitude plane should correspond to its apparent brightness. 

When computing evolutionary models, it is much easier to provide as an output the stellar equilibrium temperature than the pulsation amplitude, given that the latter can only be computed on the basis of complex non-linear pulsation models, whereas the former is directly provided as output in any standard stellar evolution code. Therefore, period shifts (with respect to some convenient reference line) {\em at fixed temperature} are routinely computed when synthetic HB models are produced (see, e.g., Lee, Demarque, \& Zinn 1990; Pritzl et al. 2002). Observers, on the other hand, generally compute period shifts {\em at fixed amplitude}. If amplitude is a good indicator of temperature, then the two diagrams are fairly equivalent, and the period shifts computed by modelers and observers can be directly compared.

V171 with a period of about 0.55~d and an amplitude of only 0.84~mag in $B$ occupies an unusual position in Figure 5. A possible explanation is that, as its period might suggest, it is a double-mode variable (0.55~d is the approximate fundamental period for M15 double-mode pulsators). A period of 0.3547~d does phase the data well, but this period is not at the proper ratio for the first-overtone period. Also, the plot at this period strongly resembles that of a fundamental pulsator. If V171 is, in fact, a double-mode pulsator, it appears as if the primary mode of pulsation is the fundamental mode, which would be unusual. Another possibility is that it is a Blazhko variable pulsating at less than maximum amplitude as discussed below.

In general, the expectation that the brighter RR Lyraes should fall closer to the OoII line in the Bailey diagram and that the dimmer ones should fall closer to the OoI line is confirmed in Figure 5. The 9 variables closest to the OoI line have $\langle B\rangle = 16.262$, although there are 2 anomalies, V77 at $\langle B\rangle = 16.118$ and V163 at $\langle B\rangle = 16.043$, the brightest RRab in our data. The 11 variables closest to the OoII line have $\langle B\rangle = 16.182$. There are 7 variables that lie well to the right of the OoII line in Figure 5. They are V9, V19, V20, V47, V49, V55, and V162. V9, V19, V47, and V162 are relative bright with $\langle B\rangle = 16.168$. However, V20, V49, and V55 are relative faint with $\langle B\rangle = 16.277$.  

There are a number of explanations for why the luminosity indicated by the position of a variable on the Bailey diagram does not correspond to its apparent magnitude. As discussed above, if the star is a blended star, it will appear brighter than its pulsational properties indicate. If the star is not a member of the cluster, but is at a greater or lesser distance, its pulsational properties will not correspond to its apparent brightness. Ratnatunga \& Bahcall (1985) indicate an expected 0.1 field stars with $B-V$ < 0.8 per square arcminute near the horizontal branch. The ``Rozhen'' images cover an area of about 32 square arcminutes, so we might expect 3 or 4 interloping field stars.

Another factor that might result in the location of the variable on the Bailey diagram not correctly representing is apparent brightness is the Blazhko effect. For the RRab variables in Figures 6 and 7, there is considerable scatter. 
Clement \& Shelton (1999) pointed out that Blazhko variables introduce
scatter into 
the period-amplitude diagram. This is because the Clement \& Shelton period-amplitude 
relations for OoI and OoII variables only apply when they are at maximum amplitude. 
The scatter is introduced when the light curve of a Blazhko variable represents it at less than its maximum amplitude. Additional intrinsic scatter can also be present as a consequence of RR Lyrae stars being in different evolutionary stages (see Pritzl et al. 2002 for a discussion).

\section{Notes On Individual Stars}

V39 --- As seen from the light curve in Figure~1, 0.38955~d does not phase the data well. In M15 double-mode pulsators have first-overtone periods of about 0.4~d and fundamental-mode periods of about 0.55~d. Thus,
V39 may be a double-mode pulsator. There have been many variables previously identified as double-mode pulsators and many new candidates in our data. However, our data are not adequate to firmly constrain secondary periods for suspected double-mode RR Lyrae. 

V139 --- This variable and ZK44 are separated by only 1.15 arcsec. 
Table~1 assumes V139 and ZK44 are the same variable, though this might not be the 
case. Our period differs from the one in OT03. However, our data are limited and it 
is not clear that our period is correct. 

V159 --- In OT03 this variable is listed as a combination of three variables.

V160 --- This variable and ZK52 are separated by only 0.37 arcsec. 
Table~1 assumes that these are the same variable.

V161 --- This variable and ZK39 are separated by only 0.05 arcsec. 
Table~1 assumes that these are the same variable. 

V162 --- This variable and ZK18 are separated by only 0.71 arcsec. 
Table~1 assumes that these are the same variable. 

V164 --- This variable and ZK37 are separated by only 0.11 arcsec. 
Table~1 assumes that these are the same variable. OT03 give a period of 
0.4275~d while ZK03 suggest that ZK37 may be a second 
overtone pulsator. The star we located is 0.66 arcsec from ZK37 and 0.75~arcsec 
from V164. Although our star appears to be variable, we could not determine a period that phased the data well.

V165 --- This variable and ZK32 are separated by only 0.03 arcsec.  
Table~1 assumes that these are the same variable. OT03 give a 
period of 0.4339~d while ZK03 do not attempt to identify 
the variable type for ZK32. The variable we located is 0.11~arcsec from V165 and 
0.18 arcsec from ZK32. Our period differs from the one in OT03. However, our data are 
limited 
and it is not clear that our period is correct. 

V166 --- This variable and ZK22 are separated by only 0.49 arcsec.  
Table~1 assumes that these are the same variable.

V169 --- This variable and ZK23 are separated by only 0.41 arcsec.  
Table~1 assumes that these are the same variable.

V172 --- This variable and ZK13 are separated by only 0.93 arcsec.  
Table~1 assumes that these are the same variable.

V173 --- This variable and ZK11 are separated by only 0.96 arcsec.  
Table~1 assumes that these are the same variable.

V174 --- This variable and ZK10 are separated by only 0.93 arcsec.  
Table~1 assumes that these are the same variable. 

V175 --- This variable and ZK6 are separated by only 1.19 arcsec. 
Table~1 assumes that these are the same variable.

V177 --- This variable and ZK64 are separated by only 0.66 arcsec. 
Table~1 assumes that these are the same variable.

V178 --- This variable and ZK55 are separated by only 0.59 arcsec. 
Table~1 assumes that these are the same variable. OT03 give a period of 
0.2860~d while ZK03 suggest that ZK37 may be a second-overtone pulsator. 
However, the OT03 period does not phase our data well. 
Our best period is 0.3983~d, with the light curve suggesting a double-mode pulsator.

V181 --- This variable and ZK14 are separated by only 0.63 arcsec.  
Table~1 assumes that these are the same variable. OT03 give a period of 
0.5868~d. Our best period is 0.40097~d.These periods suggest that V181 is a double-mode pulsator.

ZK34 --- ZK03 suggest that this variable may be a second overtone 
pulsator. Our period of 0.29765~d could be consistent with this.

ZK62 --- We did locate a star within 0.59 arcsec of this variable. Our data were very 
noisy, but did produce a slight periodic brightening at about 0.05~d.

ZK74 --- This star has a very unusual light curve for a period of 0.3195~d. ZK03 
indicate that this star may be an RRab variable. However, no longer 
period would properly phase our data.

NV14 --- Our variable is 3.4~arcsec from the Clement Catalog coordinates for V109. We assume we did not detect V109 and that our star is a new variable.

For the following stars, our data suggest variability, but no period could
 be determined from our data: V108, V109, V110, V112, V115, V116, V137, V146, V147, V148, V149, V150, V153, V164, ZK3, ZK44, ZK47, and ZK62.

\section{Summary}

In this paper, we have presented new $BVI$ CCD photometry for variable stars in the 
globular cluster M15. Our photometry was obtained with both the image subtraction package 
ISIS (Alard 2000) and with DAOPHOT/ALLFRAME (Stetson 1994). Our analysis has led to the first period determinations for 40 
previously known variables. Improved periods were also obtained for 67 previously 
known variables, and 14 new variables were discovered in the course of our study. 
Our results bring the grand total of M15 RR Lyrae variables to $N_{\rm RR} = 194$, implying 
a specific RR Lyrae fraction of 
$S_{\rm RR} = N_{\rm RR} \times 10^{0.4\, (7.5+M_V)} = 41.7$ (for an absolute magnitude of the cluster $M_V = -9.17$ Harris 1996), compared with the 
value 18.9 as given in the 2003 edition of the Harris (1996) catalog.

\acknowledgments
We warmly thank H. Markov for the observations carried out in September 2001.
Support for M.C. was provided by Proyecto FONDECYT Regular No. 1071002. HAS thanks the National Science Foundation for support under grant AST 0607249. J.B. is supported by Fondap Center for Astrophysics 15010003 and received partial support from Centro de Astrof\'{\i}sica de Valpara\'{\i}so, Universidad de Valpar\'{\i}so.

\clearpage

\begin{deluxetable}{lcclll}
\tablewidth{0pt}
\tabletypesize{\footnotesize}
\tablecaption{M15 Variables \label{tbl-1}}
\tablehead{
\colhead{Variable}  & \colhead{RA} & \colhead{Dec}  & 
\colhead{Previous period} & \colhead{Our Period} & \colhead{Type} \\
\colhead{}  & \colhead{$\rm h\ m\ s$} & \colhead{$^\circ\ '\ ''$}  & 
\colhead{$\rm (d)$} & \colhead{$\rm (d)$} & \colhead{}}
\startdata
 V1     & $21\ 29\ 50.20$  &  $12\ 10\ 26.9$  & $1.437712$  &  1.4378   &   Ceph? \\ 
 V6     & $21\ 29\ 59.92$  &  $12\ 11\ 19.8$  & $0.665967$  &  0.6660   &   RR0   \\ 
 V7     & $21\ 29\ 58.98$  &  $12\ 11\ 16.8$  & $0.367557$  &  0.367557  &   RR1   \\ 
 V8     & $21\ 29\ 58.22$  &  $12\ 12\ 10.2$  & $0.64625 $  &  0.6462   &   RR0   \\ 
 V9     & $21\ 29\ 59.24$  &  $12\ 12\ 21.9$  & $0.715295$  &  0.715295  &   RR0   \\ 
& & & & & \\[-5pt]
 V10    & $21\ 30\ 06.85$  &  $12\ 10\ 05.5$  & $0.386382$  &  0.386382  &   RR01? \\ 
 V13    & $21\ 30\ 06.95$  &  $12\ 08\ 55.3$  & $0.57491 $  &  0.57491    &   RR0   \\ 
 V16    & $21\ 30\ 05.09$  &  $12\ 12\ 13.2$  & $0.3992  $  &  0.3992   &   RR01? \\ 
 V17    & $21\ 30\ 03.95$  &  $12\ 11\ 53.6$  & $0.428907$  &  0.4294    &   RR01  \\ 
 V18    & $21\ 30\ 03.53$  &  $12\ 11\ 44.1$  & $0.367737$  &  0.367737  &   RR1   \\ 
& & & & & \\[-5pt]
 V19    & $21\ 30\ 05.81$  &  $12\ 12\ 44.2$  & $0.57228 $  &  0.5723   &   RR0   \\ 
 V20    & $21\ 30\ 03.79$  &  $12\ 09\ 53.8$  & $0.697021$  &  0.697021     &   RR0   \\ 
 V21    & $21\ 30\ 00.60$  &  $12\ 09\ 05.5$  & $0.6488  $  &  0.6476   &   RR0   \\ 
 V24    & $21\ 29\ 51.01$  &  $12\ 09\ 56.1$  & $0.369693$  &  0.3697    &   RR1   \\ 
 V32    & $21\ 29\ 54.79$  &  $12\ 11\ 50.2$  & $0.60547 $  &  0.6044   &   RR0   \\ 
 & & & & & \\[-5pt]
 V33    & $21\ 29\ 55.51$  &  $12\ 09\ 34.1$  & $0.5839452$ &  0.5839452   &   RR0   \\ 
 V34	& $21\ 29\ 54.53$  &  $12\ 09\ 07.5$  & none       &  1.1591     &  EB?    \\
 V36    & $21\ 29\ 56.42$  &  $12\ 08\ 41.6$  & $0.624151$  &  0.624151   &   RR0   \\ 
 V37    & $21\ 29\ 56.58$  &  $12\ 08\ 45.6$  & $0.28775 $  &  0.28775   &   RR1   \\ 
 V38    & $21\ 29\ 58.85$  &  $12\ 07\ 36.9$  & $0.375265$  &  0.375265  &   RR1   \\ 
& & & & & \\[-5pt]
 V39    & $21\ 29\ 59.71$  &  $12\ 07\ 58.7$  & $0.389563$  &  0.389563   &   RR01  \\ 
 V40    & $21\ 30\ 07.27$  &  $12\ 08\ 07.1$  & $0.37735 $  &  0.3777   &   RR1   \\ 
 V41    & $21\ 30\ 02.61$  &  $12\ 09\ 08.4$  & $0.382177$  &  0.382177  &   RR01  \\ 
 V44    & $21\ 30\ 04.49$  &  $12\ 10\ 07.2$  & $0.59558 $  &  0.5945   &   RR0   \\ 
 V45    & $21\ 30\ 02.83$  &  $12\ 09\ 32.1$  & $0.677399$  &  0.677399  &   RR0   \\ 
& & & & & \\[-5pt]
 V46    & $21\ 30\ 02.20$  &  $12\ 10\ 35.7$  & $0.691466$  &  0.691466  &   RR0   \\ 
 V47    & $21\ 30\ 01.28$  &  $12\ 09\ 59.6$  & $0.602799$  &  0.6875   &   RR0   \\ 
 V48    & $21\ 30\ 02.29$  &  $12\ 12\ 33.8$  & $0.364971$  &  0.3646   &   RR1   \\ 
 V49    & $21\ 30\ 00.95$  &  $12\ 12\ 49.3$  & $0.65518 $  &  0.65518   &   RR0   \\ 
 V50	& $21\ 30\ 09.44$  &  $12\ 11\ 44.0$  & $0.29806$   &  0.29806   &   RR01   \\
& & & & & \\[-5pt]
 V51    & $21\ 29\ 58.67$  &  $12\ 11\ 34.4$  & $0.396935$  &  0.396935   &   RR01  \\ 
 V53    & $21\ 29\ 52.05$  &  $12\ 08\ 11.9$  & $0.414096$  &  0.4160     &   RR01  \\ 
 V54    & $21\ 29\ 59.00$  &  $12\ 11\ 31.4$  & $0.399568$  &  0.399568  &   RR01  \\ 
 V55    & $21\ 30\ 02.76$  &  $12\ 09\ 44.7$  & $0.748596$  &  0.7486   &   RR0   \\ 
 V56    & $21\ 30\ 02.17$  &  $12\ 10\ 03.9$  & $0.5704  $  &  0.5703    &   RR0   \\ 
& & & & & \\[-5pt]
 V57    & $21\ 30\ 03.40$  &  $12\ 09\ 08.6$  & $0.537481$  &  0.3496    &   RR1   \\ 
 V58    & $21\ 29\ 54.50$  &  $12\ 10\ 11.4$  & $0.407685$  &  0.4068   &   RR01  \\ 
 V59    & $21\ 30\ 01.07$  &  $12\ 10\ 45.7$  & $0.554792$  &  0.5546   &   RR0   \\ 
 V60    & $21\ 30\ 01.93$  &  $12\ 09\ 04.3$  & $0.73124 $  &  0.7146   &   RR0   \\ 
 V61    & $21\ 29\ 53.73$  &  $12\ 09\ 21.4$  & $0.39881 $  &  0.3995   &   RR01  \\ 
& & & & & \\[-5pt]
 V62    & $21\ 29\ 53.41$  &  $12\ 10\ 41.4$  & $0.3773  $  &  0.3773    &   RR01  \\ 
 V63    & $21\ 30\ 01.59$  &  $12\ 10\ 34.2$  & $0.68332 $  &  0.6469   &   RR0   \\ 
 V64    & $21\ 29\ 55.13$  &  $12\ 10\ 22.6$  & $0.3642  $  &  0.3642   &   RR1   \\ 
 V65    & $21\ 29\ 51.33$  &  $12\ 09\ 24.2$  & $0.718196$  &  0.7182   &   RR0   \\ 
 V66    & $21\ 29\ 53.65$  &  $12\ 08\ 10.6$  & $0.37935 $  &  0.37935   &   RR1   \\ 
& & & & & \\[-5pt]
 V67    & $21\ 29\ 52.40$  &  $12\ 09\ 52.5$  & $0.404627$  &  0.4064   &   RR01? \\ 
 V68    & $21\ 29\ 56.13$  &  $12\ 10\ 15.3$  &  none     &  0.3771   &   RR1   \\ 
 V69    & $21\ 29\ 55.79$  &  $12\ 09\ 37.2$  &   none      &  0.5868   &   RR0   \\ 
 V70    & $21\ 29\ 55.98$  &  $12\ 09\ 43.6$  &   none      &  0.3676   &   RR1   \\ 
 V71    & $21\ 29\ 55.94$  &  $12\ 09\ 50.4$  &   none      &  0.3737   &   RR1   \\ 
& & & & & \\[-5pt]
 V73    & $21\ 29\ 58.04$  &  $12\ 10\ 23.2$  &   none      &  0.3113    &   RR1   \\ 
 V74    & $21\ 30\ 00.90$  &  $12\ 08\ 37.7$  & $0.29601 $  &  0.29601   &   RR1   \\ 
 V75    & $21\ 29\ 58.47$  &  $12\ 09\ 32.1$  &   none      &  0.3655   &   RR1   \\ 
 V76    & $21\ 29\ 58.37$  &  $12\ 09\ 34.3$  &   none      &  0.3841   &   RR1   \\ 
 V77    & $21\ 29\ 57.57$  &  $12\ 09\ 40.5$  &   none      &  0.5960   &   RR0   \\ 
& & & & & \\[-5pt]
 V78    & $21\ 29\ 57.83$  &  $12\ 10\ 50.6$  & $0.398879$  &  0.6648   &   RR0   \\ 
 V80    & $21\ 29\ 55.08$  &  $12\ 09\ 36.2$  &   none      &   0.6642   &   RR0   \\ 
 V81    & $21\ 29\ 56.83$  &  $12\ 09\ 57.9$  &   none      &  0.6033   &   RR0   \\ 
 V82    & $21\ 29\ 56.88$  &  $12\ 10\ 04.9$  &   none      &  0.3691   &   RR1   \\ 
 V83    & $21\ 29\ 59.39$  &  $12\ 09\ 56.8$  &   none      &  0.7263   &   RR0   \\ 
& & & & & \\[-5pt]
 V84    & $21\ 29\ 59.51$  &  $12\ 09\ 48.9$  &   none      &  0.5872   &   RR0   \\ 
 V86    & $21\ 29\ 59.18$  &  $12\ 10\ 07.5$  & $17.109  $  &  ?    &   Ceph? \\ 
 V87	& $21\ 30\ 00.03$  &  $12\ 09\ 39.3$  &   none      &  0.3208   &   RR1   \\
 V88    & $21\ 29\ 58.43$  &  $12\ 10\ 29.6$  &   none      &  0.6822   &   RR0   \\ 
 V89    & $21\ 29\ 56.70$  &  $12\ 09\ 55.8$  &   none      &  0.3789   &   RR1   \\ 
& & & & & \\[-5pt]
 V90    & $21\ 30\ 00.48$  &  $12\ 10\ 07.8$  &   none      &  0.3823   &   RR1   \\ 
 V91    & $21\ 30\ 02.89$  &  $12\ 10\ 31.1$  &   none      &  0.6235    &   RR0   \\ 
 V92    & $21\ 29\ 59.00$  &  $12\ 09\ 37.8$  &   none      &  0.4120   &   RR1   \\ 
 V93    & $21\ 30\ 00.30$  &  $12\ 09\ 29.5$  &   none      &  0.3970   &   RR01  \\ 
 V94    & $21\ 29\ 58.48$  &  $12\ 10\ 31.4$  &   none      &  0.6630     &   RR0   \\ 
& & & & & \\[-5pt]
 V97    & $21\ 29\ 52.87$  &  $12\ 10\ 31.8$  & $0.696336$  &  0.696336  &   RR0   \\ 
 V106   & $21\ 29\ 56.13$  &  $12\ 10\ 17.0$  &   none      &  0.3957    &   RR01  \\ 
 V111   & $21\ 30\ 01.20$  &  $12\ 10\ 03.3$  &   none      &  0.4089   &   RR01  \\ 
 V114   & $21\ 29\ 58.19$  &  $12\ 10\ 53.3$  &   none      &  0.3579    &   RR1   \\ 
 V117   & $21\ 29\ 59.31$  &  $12\ 09\ 27.3$  &   none      &  0.6869   &   RR0   \\ 
& & & & & \\[-5pt]
 V118   & $21\ 29\ 59.63$  &  $12\ 10\ 56.0$  &   none      &  0.3000   &   RR01  \\ 
 V119   & $21\ 29\ 59.65$  &  $12\ 10\ 07.6$  &   none      &  0.4406   &   RR1   \\ 
 V128   & $21\ 29\ 58.41$  &  $12\ 10\ 02.3$  &  $0.4034 $  &  0.3892    &   RR01? \\ 
 V129   & $21\ 29\ 57.85$  &  $12\ 09\ 49.0$  & $0.6876  $  &  0.6895   &   RR0   \\ 
 V130   & $21\ 29\ 57.87$  &  $12\ 09\ 52.3$  & $0.652   $  &  0.6456   &   RR0   \\ 
& & & & & \\[-5pt]
 V131   & $21\ 29\ 57.86$  &  $12\ 10\ 00.8$  & $0.5994  $  &  0.3774    &   RR1?  \\ 
 V132   & $21\ 29\ 57.85$  &  $12\ 10\ 03.0$  & $0.3930  $  &  0.4002    &   RR01  \\
 V134   & $21\ 29\ 58.03$  &  $12\ 09\ 59.3$  & $0.3828  $  &  0.6248    &   RR0   \\ 
 V136   & $21\ 29\ 58.18$  &  $12\ 10\ 07.3$  & $0.7069  $  &  0.7232    &   RR0   \\ 
 V137	& $21\ 29\ 58.29$  &  $12\ 10\ 05.6$  & $0.3520  $  &  0.5388    &   RR1  \\
& & & & & \\[-5pt]
 V138   & $21\ 29\ 58.48$  &  $12\ 10\ 00.9$  & $0.7143  $  &  0.7290     &   RR0   \\ 
 V139   & $21\ 29\ 58.43$  &  $12\ 10\ 11.5$  & $0.3258  $  &  0.3239    &   RR1   \\ 
 V141   & $21\ 29\ 58.97$  &  $12\ 10\ 03.6$  & $0.3644  $  &  0.3644   &   RR1   \\ 
 V142   & $21\ 29\ 58.60$  &  $12\ 10\ 01.0$  & $1.2411  $  &  1.2266    &   Ceph? \\ 
 V144   & $21\ 29\ 59.30$  &  $12\ 10\ 01.4$  & $0.299   $  &  0.2985    &   RR1   \\ 
& & & & & \\[-5pt]
 V145   & $21\ 29\ 59.34$  &  $12\ 10\ 03.2$  & $0.4224  $  &  0.4182    &   RR01? \\ 
 V159   & $21\ 29\ 58.35$  &  $12\ 10\ 02.6$  & $0.6385  $  &  0.6933    &   RR0   \\ 
 V160   & $21\ 29\ 58.10$  &  $12\ 09\ 48.6$  & $0.3529  $  &  0.3531    &   RR1   \\ 
 V161   & $21\ 29\ 58.51$  &  $12\ 09\ 49.5$  & $1.2773  $  &  0.5573    &   RR0   \\ 
 V162   & $21\ 29\ 59.55$  &  $12\ 09\ 50.0$  & $0.7299  $  &  0.7337    &   RR0   \\ 
& & & & & \\[-5pt]
 V163   & $21\ 29\ 59.25$  &  $12\ 09\ 52.3$  & $0.5601  $  &  0.5538    &   RR0   \\ 
 V165   & $21\ 29\ 58.79$  &  $12\ 10\ 03.7$  & $0.4339  $  &  0.6251   &   RR0   \\ 
 V166   & $21\ 29\ 59.38$  &  $12\ 10\ 01.1$  & $0.3669  $  &  0.3656   &   RR1   \\ 
 V167   & $21\ 29\ 59.83$  &  $12\ 10\ 03.7$  & $0.47    $  &  0.3192   &   RR1   \\ 
 V168   & $21\ 29\ 59.72$  &  $12\ 10\ 05.8$  &  $0.7225$   &  0.6029   &   RR0   \\ 
& & & & & \\[-5pt]
 V169   & $21\ 29\ 59.36$  &  $12\ 10\ 09.0$  & $0.5555  $  &  0.5549   &   RR0   \\ 
 V170   & $21\ 29\ 58.81$  &  $12\ 09\ 21.9$  & $0.6817  $  &  0.6781    &   RR0   \\ 
 V171   & $21\ 29\ 59.82$  &  $12\ 09\ 40.0$  & $0.4578$    &  0.5527    &   RR0   \\ 
 V172   & $21\ 29\ 59.94$  &  $12\ 09\ 49.1$  & $0.5443  $  &  0.4317    &   RR01   \\
 V173   & $21\ 30\ 00.06$  &  $12\ 10\ 01.8$  & $0.3747  $  &  0.3784    &   RR1   \\ 
& & & & & \\[-5pt]
 V174   & $21\ 30\ 00.20$  &  $12\ 10\ 13.5$  & $0.2821  $  &  0.4006   &   RR01  \\ 
 V175   & $21\ 30\ 00.48$  &  $12\ 10\ 13.6$  & $0.4282  $  &  0.7540     &   RR0   \\ 
 V177   & $21\ 29\ 57.67$  &  $12\ 10\ 14.9$  & $0.407   $  &  0.4116   &   RR01  \\ 
 V178   & $21\ 29\ 58.02$  &  $12\ 10\ 19.3$  & $0.286   $  &  0.4009   &   RR01  \\ 
 V181   & $21\ 29\ 59.81$  &  $12\ 10\ 25.1$  & $0.5868  $  &  0.2232   &   RR01  \\ 
 & & & & & \\[-5pt]
 ZK3	& $21\ 30\ 01.17$  &  $12\ 10\ 11.0$  & none      &  0.6310   &   RR0   \\
 ZK4    & $21\ 30\ 00.91$  &  $12\ 09\ 56.1$  &   none      &  0.3829   &   RR1   \\ 
 ZK5    & $21\ 30\ 00.87$  &  $12\ 09\ 56.8$  &   none      &  0.3647   &   RR1   \\ 
 ZK6    &  \multicolumn{ 2}{c} {=V 175}       &             &            &   RR0   \\ 
 ZK10   &  \multicolumn{ 2}{c} {=V 174}       &             &            &   RR1   \\ 
& & & & & \\[-5pt]
 ZK11   &  \multicolumn{ 2}{c} {=V 173}       &             &            &   RR1   \\ 
 ZK13   &  \multicolumn{ 2}{c} {=V 172}       &             &            &   RR1   \\ 
 ZK14   &  \multicolumn{ 2}{c} {=V 181}       &             &            &   RR0   \\ 
 ZK18   &  \multicolumn{ 2}{c} {=V 162}       &             &            &   RR0   \\ 
 ZK22   &  \multicolumn{ 2}{c} {=V 166}       &             &            &   RR1   \\ 
& & & & & \\[-5pt]
 ZK23   &  \multicolumn{ 2}{c} {=V 169}       &             &            &   RR0   \\  
 ZK32   &  \multicolumn{ 2}{c} {=V 165}       &             &            &   RR0   \\ 
 ZK34   & $21\ 29\ 58.75$  &  $12\ 09\ 28.7$  &   none      &  0.2977   &   RR1?  \\ 
 ZK39   &  \multicolumn{ 2}{c} {=V 161}       &             &            &   RR0   \\ 
 ZK44   &  \multicolumn{ 2}{c} {=V 139}       &             &            &   RR1   \\ 
& & & & & \\[-5pt]
 ZK52   &  \multicolumn{ 2}{c} {=V 160}       &             &            &   RR1   \\ 
 ZK55   &  \multicolumn{ 2}{c} {=V 178}       &             &            &   RR1   \\ 
 ZK63   & $21\ 29\ 57.81$  &  $12\ 10\ 21.0$  &   none      &   0.4020  &   RR01  \\ 
 ZK64   &  \multicolumn{ 2}{c} {=V 177}       &             &            &   RR1   \\ 
 ZK67   & $21\ 29\ 57.45$  &  $12\ 09\ 43.6$  &   none      &   0.5709  &   RR0   \\ 
& & & & & \\[-5pt]
 ZK69   & $21\ 29\ 57.13$  &  $12\ 10\ 21.4$  &   none      &   0.4168  &   RR01  \\ 
 ZK74   & $21\ 29\ 56.25$  &  $12\ 10\ 30.9$  &   none      &   0.3197  &   RR1   \\ 
 ZK78   & $21\ 29\ 55.90$  &  $12\ 10\ 27.8$  &   none      &   0.4256  &   RR1   \\ 
 ZK80   & $21\ 29\ 55.86$  &  $12\ 10\ 09.2$  &   none      &   0.3873  &   RR1   \\ 
 NV1    & $21\ 29\ 53.93$  &  $12\ 11\ 13.6$  &   none      &   0.0585  &   SX Phe\\ 
& & & & & \\[-5pt]
 NV2    & $21\ 29\ 56.22$  &  $12\ 09\ 37.9$  &   none      &   0.4166  &   RR1?  \\ 
 NV3    & $21\ 29\ 58.50$  &  $12\ 10\ 40.2$  &   none      &   0.4210  &   RR01?  \\ 
 NV4    & $21\ 29\ 58.89$  &  $12\ 09\ 02.7$  &   none      &   0.3776  &   RR1   \\ 
 NV5    & $21\ 29\ 59.42$  &  $12\ 09\ 13.4$  &   none      &   0.6101   &   RR0   \\ 
 NV6    & $21\ 30\ 02.69$  &  $12\ 10\ 31.9$  &   none      &   0.3909  &   RR1   \\ 
& & & & & \\[-5pt]
 NV7    & $21\ 30\ 00.40$  &  $12\ 09\ 57.8$  &   none      &   0.3854   &   RR1   \\ 
 NV8    & $21\ 30\ 00.84$  &  $12\ 09\ 43.4$  &   none      &   0.6294  &   RR0   \\ 
 NV9    & $21\ 30\ 02.93$  &  $12\ 10\ 09.5$  &   none      &   0.3915  &   RR1   \\ 
 NV10   & $21\ 30\ 03.13$  &  $12\ 10\ 32.4$  &   none      &   0.7594  &   RR0   \\ 
 NV11   & $21\ 29\ 59.55$  &  $12\ 09\ 23.6$  &   none      &   0.2957  &   RR01  \\ 
& & & & & \\[-5pt]
 NV12   & $21\ 29\ 59.31$  &  $12\ 09\ 24.3$  &   none      &   0.4104  &   RR01  \\ 
 NV13   & $21\ 30\ 00.03$  &  $12\ 09\ 39.3$  &   none      &   0.3208  &   ?     \\ 
 NV14   & $21\ 29\ 59.32$  &  $12\ 09\ 29.1$  &   none      &   0.2993  &   RR1   \\ 
\enddata
\end{deluxetable}

\clearpage

\begin{deluxetable}{rrrrrcc}
\tablewidth{0pt}
\tabletypesize{\footnotesize}
\tablecaption{M15 Variables: Positions Relative to the Cluster Center \label{tbl
-2}}
\tablecolumns{7}
\tablehead{
\colhead{Variable}  & \multicolumn{1}{c}{prev $x$}  & \multicolumn{1}{c}{prev $y
$}  & \multicolumn{1}{c}{our $x$}   & \multicolumn{1}{c}{our $y$}    & \colhead{
comment} & \colhead{difference} \\
 \colhead{} & \multicolumn{1}{c}{(arcsec)} & \multicolumn{1}{c}{(arcsec)} 
 & \multicolumn{1}{c}{(arcsec)} & \multicolumn{1}{c}{(arcsec)} &           
 \colhead{}  & \colhead{(arcsec)}} 
\startdata
V11   &     $     172.30$  &  $- 21.80$ &  $ 171.67$ & $-21.43$   &          &     $0.73$ \\
V37   &     $-     25.20$  &  $- 77.40$ &  $- 25.69$ & $-77.27$   &          &     $0.51$ \\
V39   &     $      20.50$  &  $-124.80$ &  $  20.18$ & $-124.45$  &          &     $0.47$ \\
V47   &     $      45.70$  &  $-  4.30$ &  $  43.49$ & $- 3.78$   &          &     $2.27$ \\
V53   &     $-     92.60$  &  $-111.00$ &  $- 92.29$ & $-110.50$  &          &     $0.59$ \\
 & & & & & & \\[-5pt]
V58   &     $-     55.60$  &  $   8.80$ &  $- 56.06$ & $  8.68$   &          &     $0.48$ \\
V60   &     $      53.40$  &  $- 59.30$ &  $  52.92$ & $-59.21$   &          &     $0.49$ \\
V61   &     $-     67.30$  &  $- 40.20$ &  $- 67.47$ & $-41.21$   &          &     $1.03$ \\
V63   &     $      49.80$  &  $  31.00$ &  $  48.11$ & $ 30.76$   &          &     $1.71$ \\
V68   &     $-     31.80$  &  $  12.60$ &  $- 32.11$ & $ 12.41$   &          &     $0.36$ \\
 & & & & &  & \\[-5pt]
V69   &     $-     37.00$  &  $- 25.20$ &  $- 37.19$ & $-25.63$   &          &     $0.47$ \\
V70   &     $-     34.00$  &  $- 19.20$ &  $- 34.38$ & $-19.25$   &          &     $0.38$ \\
V71   &     $-     34.80$  &  $- 12.60$ &  $- 34.96$ & $-12.45$   &          &     $0.22$ \\
V73   &     $-      3.70$  &  $  20.00$ &  $-  4.05$ & $ 20.12$   &      &     $0.37$ \\
V75   &     $       2.20$  &  $- 30.30$ &  $   2.16$ & $-30.99$   &          &     $0.69$ \\
 & & & & &  & \\[-5pt]
V76   &     $       0.70$  &  $- 28.90$ &  $   0.70$ & $-28.78$   &          &     $0.12$ \\
V77   &     $-     11.80$  &  $- 22.90$ &  $- 11.04$ & $-22.51$   &          &     $0.85$ \\
V80   &     $-     47.40$  &  $- 26.60$ &  $- 47.61$ & $-26.56$   &          &     $0.21$ \\
V81   &     $-     21.50$  &  $-  5.90$ &  $- 21.87$ & $- 5.04$   &          &     $0.94$ \\
V82   &     $-     20.10$  &  $   2.40$ &  $- 21.12$ & $  1.95$   &          &     $1.11$ \\
V83   &     $      16.30$  &  $-  7.40$ &  $  15.73$ & $- 6.40$   &          &     $1.15$ \\
& & & & & \\[-5pt]
V84   &     $      17.53$  &  $- 14.20$ &  $  17.47$ & $-14.30$   &          &     $0.12$ \\
V86   &     $      12.60$  &  $   4.40$ &  $  12.66$ & $  4.32$   &          &     $0.10$ \\
V88   &     $       2.20$  &  $  26.60$ &  $   1.69$ & $ 26.48$   &      &     $0.52$ \\
V89   &     $-     23.70$  &  $-  6.70$ &  $- 23.78$ & $- 7.13$   &          &     $0.44$ \\
V90   &     $      31.10$  &  $   4.40$ &  $  31.76$ & $  4.49$   &          &     $0.67$ \\
& & & & & \\[-5pt]
V91   &     $      67.30$  &  $  28.90$ &  $  67.20$ & $ 27.54$   &          &     $1.36$ \\
V92   &     $       9.60$  &  $- 25.20$ &  $   9.96$ & $-25.34$   &          &     $0.39$ \\
V93   &     $      27.40$  &  $- 33.30$ &  $  29.03$ & $-33.77$   &          &     $1.70$ \\
V94   &     $       3.70$  &  $  28.90$ &  $   2.43$ & $ 28.27$   &          &     $1.42$ \\
V97   &     $-     79.50$  &  $  29.30$ &  $- 79.96$ & $ 29.22$   &          &     $0.47$ \\
& & & & & \\[-5pt]
V106   &     $-     30.30$  &  $  12.80$ &  $- 32.11$ & $ 13.91$   &          &     $2.12$ \\
V111   &     $      41.70$  &  $-  0.70$ &  $  42.32$ & $- 0.08$   &          &     $0.88$ \\
V114   &     $       0.91$  &  $  49.93$ &  $-  1.78$ & $ 49.88$   &          &     $2.69$ \\
V117   &     $      15.84$  &  $- 36.80$ &  $  14.49$ & $-35.87$   &          &     $1.64$ \\
V118   &     $      17.60$  &  $  55.67$ &  $  19.37$ & $ 52.74$   &          &     $3.42$ \\
& & & & & \\[-5pt]
V119   &     $      20.74$  &  $   3.51$ &  $  19.57$ & $  4.37$   &          &     $1.45$ \\
V128   &     $       1.03$  &  $-  0.86$ &  $-  1.00$ & $  1.40$   &          &     $3.04$ \\
V129   &     $-      7.28$  &  $- 14.02$ &  $-  6.88$ & $-14.00$   &          &     $0.40$ \\
V130   &     $-      6.95$  &  $ -10.71$ &  $-  6.11$ & $-10.68$   &          &     $0.84$ \\
V131   &     $-      7.05$  &  $-  2.23$ &  $-  6.71$ & $- 2.21$   &      &     $0.34$ \\
& & & & & \\[-5pt]
V134   &     $-      4.44$  &  $-  3.79$ &  $-  4.30$ & $- 3.79$   &          &     $0.14$ \\
V136   &     $-      2.24$  &  $   4.20$ &  $-  1.36$ & $  4.49$   &          &     $0.93$ \\
V137   &     $-      0.71$  &  $   2.52$ &  $-  0.58$ & $  2.54$   &          &     $0.13$ \\
V138   &     $       3.07$  &  $-  4.12$ &  $   3.58$ & $- 4.35$   &          &     $0.56$ \\
V139   &     $       1.43$  &  $   8.38$ &  $   0.98$ & $  7.54$   &  ZK44    &     $0.95$ \\
& & & & & \\[-5pt]
V141   &     $       9.23$  &  $   0.41$ &  $   9.03$ & $  0.49$   &          &     $0.22$ \\
V142   &     $       3.86$  &  $-  2.11$ &  $   4.37$ & $- 2.15$   &          &     $0.51$ \\
V144   &     $      13.89$  &  $-  1.81$ &  $  14.13$ & $- 1.76$   &          &     $0.25$ \\
V145   &     $      14.38$  &  $   0.31$ &  $  15.03$ & $  0.32$   &          &     $0.65$ \\
V155   &     $-     10.64$  &  $   7.23$ &  $- 10.58$ & $  7.25$   &          &     $0.06$ \\
& & & & & \\[-5pt]
V159   &     $       0.92$  &  $-  0.23$ &  $   0.46$ & $- 0.50$   &          &     $0.53$ \\
V160   &     $-      3.85$  &  $- 14.43$ &  $-  3.24$ & $-14.46$   &  ZK52    &     $0.61$ \\
V161   &     $       2.89$  &  $- 13.45$ &  $   2.79$ & $- 13.6$   &  ZK39    &     $0.18$ \\
V162   &     $      18.39$  &  $- 13.00$ &  $  18.06$ & $-13.21$   &  ZK18    &     $0.39$ \\
V163   &     $      13.63$  &  $- 10.70$ &  $  13.66$ & $-10.88$   &          &     $0.18$ \\
& & & & & \\[-5pt]
V164   &     $       3.05$  &  $   3.70$ &  $   2.82$ & $  3.48$   &  ZK37    &     $0.32$ \\
V165   &     $       6.73$  &  $   0.69$ &  $   6.93$ & $  0.56$   &  ZK32    &     $0.24$ \\
V166   &     $      15.78$  &  $-  2.01$ &  $  15.59$ & $- 2.10$   &  ZK22    &     $0.21$ \\
V167   &     $      22.68$  &  $   0.70$ &  $  22.20$ & $  0.46$   &          &     $0.54$ \\
V168   &     $      20.99$  &  $   2.74$ &  $  20.59$ & $  2.57$   &       &     $0.43$ \\
& & & & & \\[-5pt]
V169   &     $      15.47$  &  $   5.97$ &  $  15.31$ & $  5.80$   &  ZK23    &     $0.23$ \\
V170   &     $       7.04$  &  $- 41.27$ &  $   7.14$ & $-41.21$   &          &     $0.12$ \\
V171   &     $      22.53$  &  $- 23.08$ &  $  22.01$ & $-23.03$   &       &     $0.52$ \\
V172   &     $      24.21$  &  $- 13.99$ &  $  23.64$ & $-15.24$   &  ZK13    &     $1.37$ \\
V173   &     $      26.21$  &  $-  1.19$ &  $  25.58$ & $- 1.47$   &  ZK11    &     $0.69$ \\
& & & & & \\[-5pt]
V174   &     $      28.20$  &  $  10.52$ &  $  27.66$ & $ 10.21$   &  ZK10    &     $0.62$ \\
V175   &     $      32.65$  &  $  10.62$ &  $  31.77$ & $ 10.28$   &  ZK6     &     $0.94$ \\
V177   &     $-     10.45$  &  $  11.80$ &  $-  9.50$ & $ 11.86$   &  ZK64    &     $0.95$ \\
V178   &     $-      5.08$  &  $  16.19$ &  $-  4.35$ & $ 16.22$   &  ZK55    &     $0.73$ \\
V181   &     $      22.37$  &  $  22.11$ &  $  21.95$ & $ 21.84$   &  ZK14    &     $0.50$ \\
& & & & & \\[-5pt]
ZK3   &     $      41.65$  &  $   7.61$ &  $  41.90$ & $  7.62$   &          &     $0.25$ \\
ZK4   &     $      37.66$  &  $-  7.09$ &  $  38.05$ & $- 7.25$   &          &     $0.42$ \\
ZK5   &     $      37.26$  &  $-  6.84$ &  $  37.46$ & $- 6.54$   &          &     $0.36$ \\
ZK6   &     $      31.51$  &  $  10.34$ &  $  31.77$ & $ 10.28$   &  V175    &     $0.27$ \\
ZK10  &     $      27.30$  &  $  10.25$ &  $  27.66$ & $ 10.21$   &  V174    &     $0.36$ \\
& & & & & \\[-5pt]
ZK11  &     $      25.27$  &  $-  1.47$ &  $  25.58$ & $- 1.47$   &  V173    &     $0.31$ \\
ZK13  &     $      23.29$  &  $- 14.21$ &  $  23.64$ & $-15.44$   &  V172    &     $1.28$ \\
ZK14  &     $      21.82$  &  $  21.92$ &  $  21.95$ & $ 21.84$   &  V181    &     $0.15$ \\
ZK18  &     $      17.70$  &  $- 13.16$ &  $  18.06$ & $-13.21$   &  V162    &     $0.36$ \\
ZK22  &     $      15.27$  &  $-  2.06$ &  $  15.59$ & $- 2.10$   &  V166    &     $0.32$ \\
& & & & & \\[-5pt]
ZK23  &     $      15.07$  &  $   5.78$ &  $  15.31$ & $  5.80$   &  V169    &     $0.24$ \\
ZK32  &     $       6.65$  &  $   0.70$ &  $   6.93$ & $  0.56$   &  V165    &     $0.31$ \\
ZK34  &     $       5.85$  &  $- 34.44$ &  $   6.27$ & $-34.41$   &          &     $0.42$ \\
ZK37  &     $       3.12$  &  $   3.63$ &  $   2.82$ & $  3.48$   &  V164    &     $0.34$ \\
ZK39  &     $       2.91$  &  $- 13.46$ &  $   2.79$ & $-13.60$   &  V161    &     $0.18$ \\
& & & & & \\[-5pt]
ZK44  &     $       1.21$  &  $   7.25$ &  $   1.51$ & $  7.09$   &  V139    &     $0.34$ \\
ZK47  &     $       0.81$  &  $-  1.94$ &  $   0.31$ & $- 1.90$   &          &     $0.50$ \\
ZK52  &     $-      3.54$  &  $- 14.54$ &  $-  3.24$ & $-14.46$   &  V160    &     $0.31$ \\
ZK55  &     $-      4.49$  &  $  16.33$ &  $-  4.35$ & $ 16.22$   &  V178    &     $0.18$ \\
ZK62  &     $-      7.39$  &  $-  8.70$ &  $-  7.04$ & $- 8.64$   &  Phe     &     $0.36$ \\
& & & & & \\[-5pt]
ZK63  &     $-      7.68$  &  $  18.19$ &  $-  7.43$ & $ 17.94$   &          &     $0.35$ \\
ZK64  &     $-      9.75$  &  $  11.73$ &  $-  9.50$ & $ 11.86$   &  V177    &     $0.28$ \\
ZK67  &     $-     13.04$  &  $- 19.32$ &  $- 12.79$ & $-19.40$   &          &     $0.26$ \\
ZK69  &     $-     17.58$  &  $  18.47$ &  $- 17.41$ & $ 18.41$   &          &     $0.18$ \\
ZK74  &     $-     30.51$  &  $  28.17$ &  $- 30.32$ & $ 27.99$   &          &     $0.26$ \\
& & & & & \\[-5pt]
ZK78  &     $-     35.53$  &  $  25.13$ &  $- 35.47$ & $ 24.93$   &          &     $0.21$ \\
ZK80  &     $-     36.27$  &  $   6.46$ &  $- 36.09$ & $  6.34$   &          &     $0.22$ 
\enddata             
\end{deluxetable}

\clearpage

\tabcolsep=5pt

\begin{deluxetable}{lccccccccc}
\tablecaption{M15 Variables \label{tbl-3}}
\tablewidth{0pt}
\tabletypesize{\footnotesize}
\tablehead{
\colhead{Variable}  & \colhead{$\langle V \rangle_{\rm mag}$} & \colhead{$\langle V \rangle_{\rm int}$}  & 
\colhead{$A_V$} & \colhead{$\langle B \rangle_{\rm mag}$} & \colhead{$\langle B \rangle_{\rm int}$} & \colhead{$A_B$} 
& \colhead{$\langle I \rangle_{\rm mag}$} & \colhead{$\langle I \rangle_{\rm int}$} & \colhead{$A_I$}  }
\startdata
V1            &  $15.009$  &   $14.954$   &   $0.99$   &  $15.412$  &   $15.317$   &   $1.26$   &  $14.385$  &   $14.362$  &   $0.69$       \\
V6            &  $15.797$  &   $15.757$   &   $0.94$   &  $16.224$  &   $16.180$   &   $1.14$   &  $15.156$  &   $15.142$  &   $0.54$       \\
V7            &  $15.770$  &   $15.754$   &   $0.56$   &  $16.180$  &   $16.157$   &   $0.65$   &  $15.316$  &   $15.306$  &   $0.32$       \\
V8            &  $15.795$  &   $15.763$   &   $0.89$   &  $16.241$  &   $16.190$   &   $1.05$   &  $15.178$  &   $15.163$  &   $0.51$       \\
V9            &  $15.711$  &   $15.685$   &   $0.84$   &  $16.110$  &   $16.068$   &   $1.05$   &  $15.137$  &   $15.126$  &   $0.79$       \\
 & & & & & & & & & \\[-5pt]
V10           &  $15.869$  &   $15.856$   &   $0.50$   &  $16.214$  &   $16.191$   &   $0.65$   &  $15.413$  &   $15.407$  &   $0.36$       \\
V13           &  $15.978$  &   $15.944$   &   $1.01$   &  $16.384$  &   $16.323$   &   $1.18$   &  $15.386$  &   $15.372$  &   $0.65$       \\
V16           &  $15.854$  &   $15.843$   &   $0.36$   &  $16.186$  &   $16.169$   &   $0.47$   &  $15.356$  &   $15.352$  &   $0.23$       \\
V17           &  $15.753$  &   $15.743$   &   $    $   &  $16.104$  &   $16.086$   &   $    $   &  $15.274$  &   $15.270$  &   $    $       \\
V18           &  $15.846$  &   $15.829$   &   $0.55$   &  $16.185$  &   $16.157$   &   $0.70$   &  $15.397$  &   $15.391$  &   $0.34$       \\
 & & & & & & & & & \\[-5pt]
V19           &  $15.812$  &   $15.733$   &   $1.38$   &  $16.173$  &   $16.052$   &   $1.69$   &  $15.319$  &   $15.291$  &   $0.90$       \\
V20           &  $15.779$  &   $15.751$   &   $0.90$   &  $16.266$  &   $16.220$   &   $1.09$   &  $15.137$  &   $15.129$  &   $0.42$       \\
V21           &  $15.705$  &   $15.669$   &   $0.85$   &  $16.139$  &   $16.073$   &   $1.20$   &  $15.025$  &   $15.012$  &   $0.49$       \\
V24           &  $15.877$  &   $15.859$   &   $0.50$   &  $16.216$  &   $16.188$   &   $0.78$   &  $15.341$  &   $15.336$  &   $0.37$       \\
V32           &  $15.774$  &   $15.735$   &   $0.97$   &  $16.194$  &   $16.131$   &   $1.22$   &  $15.205$  &   $15.192$  &   $0.64$       \\
 & & & & & & & & & \\[-5pt]
V33           &  $15.776$  &   $15.732$   &   $1.05$   &  $16.180$  &   $16.105$   &   $1.31$   &  $15.196$  &   $15.178$  &   $0.68$       \\
V36           &  $15.853$  &   $15.817$   &   $0.93$   &  $16.268$  &   $16.202$   &   $1.15$   &  $15.208$  &   $15.195$  &   $0.59$       \\
V37           &  $15.885$  &   $15.871$   &   $0.53$   &  $16.180$  &   $16.157$   &   $0.69$   &  $15.474$  &   $15.467$  &   $0.33$       \\
V38           &  $15.839$  &   $15.820$   &   $0.55$   &  $16.200$  &   $16.173$   &   $0.73$   &  $15.338$  &   $15.334$  &   $0.29$       \\
V39           &  $15.892$  &   $15.880$   &   $    $   &  $16.274$  &   $16.258$   &   $    $   &  $15.397$  &   $15.390$  &   $    $       \\
 & & & & & & & & & \\[-5pt]
V40           &  $15.885$  &   $15.867$   &   $    $   &  $16.213$  &   $16.185$   &   $    $   &  $15.346$  &   $15.336$  &   $    $       \\
V41           &  $15.715$  &   $15.703$   &   $    $   &  $16.090$  &   $16.074$   &   $    $   &  $15.141$  &   $15.135$  &   $    $       \\
V44           &  $15.800$  &   $15.773$   &   $0.81$   &  $16.253$  &   $16.206$   &   $1.00$   &  $15.252$  &   $15.242$  &   $0.58$       \\
V45           &  $15.694$  &   $15.651$   &   $0.91$   &  $16.166$  &   $16.099$   &   $1.12$   &  $15.089$  &   $15.076$  &   $0.53$       \\
V46           &  $15.645$  &   $15.619$   &   $0.74$   &  $16.141$  &   $16.102$   &   $0.90$   &  $15.028$  &   $15.019$  &   $0.46$       \\
 & & & & & & & & & \\[-5pt]
V47           &  $15.711$  &   $15.676$   &   $0.96$   &  $16.177$  &   $16.126$   &   $1.18$   &  $15.029$  &   $15.017$  &   $0.63$       \\
V48           &  $15.887$  &   $15.870$   &   $0.63$   &  $16.193$  &   $16.162$   &   $0.70$   &  $15.418$  &   $15.411$  &   $0.38$       \\
V49           &  $15.896$  &   $15.872$   &   $0.81$   &  $16.289$  &   $16.243$   &   $1.07$   &  $15.245$  &   $15.241$  &   $0.56$       \\
V50           &  $15.945$  &   $15.927$   &   $    $   &  $16.171$  &   $16.134$   &   $    $   &  $15.563$  &   $15.553$  &   $    $       \\
V51           &  $15.827$  &   $15.816$   &   $0.45$   &  $16.192$  &   $16.176$   &   $0.58$   &  $15.334$  &   $15.329$  &   $0.33$       \\
 & & & & & & & & & \\[-5pt]
V53           &  $15.828$  &   $15.822$   &   $    $   &  $16.213$  &   $16.203$   &   $    $   &  $15.310$  &   $15.305$  &   $    $       \\
V54           &  $15.745$  &   $15.734$   &   $0.48$   &  $16.159$  &   $16.143$   &   $0.59$   &  $15.255$  &   $15.251$  &   $0.33$       \\
V55           &  $15.746$  &   $15.732$   &   $0.67$   &  $16.277$  &   $16.254$   &   $0.75$   &  $15.076$  &   $15.071$  &   $0.39$       \\
V56           &  $15.780$  &   $15.741$   &   $1.02$   &  $16.249$  &   $16.180$   &   $1.21$   &  $15.156$  &   $15.142$  &   $0.60$       \\
V57           &  $15.793$  &   $15.773$   &   $0.58$   &  $16.142$  &   $16.113$   &   $0.69$   &  $15.350$  &   $15.343$  &   $0.36$       \\
 & & & & & & & & & \\[-5pt]
V58           &  $15.830$  &   $15.825$   &   $    $   &  $16.214$  &   $16.206$   &   $    $   &  $15.152$  &   $15.145$  &   $    $       \\
V59           &  $15.855$  &   $15.797$   &   $1.13$   &  $16.274$  &   $16.174$   &   $1.44$   &  $15.327$  &   $15.307$  &   $0.70$       \\
V60           &  $15.612$  &   $15.598$   &   $0.61$   &  $16.159$  &   $16.137$   &   $0.73$   &  $14.927$  &   $15.921$  &   $0.46$       \\
V61           &  $15.802$  &   $15.794$   &   $    $   &  $16.236$  &   $16.225$   &   $    $   &  $15.205$  &   $15.203$  &   $    $       \\
V62           &  $15.741$  &   $15.726$   &   $0.53$   &  $16.062$  &   $16.042$   &   $0.61$   &  $15.210$  &   $15.205$  &   $0.30$       \\
 & & & & & & & & & \\[-5pt]
V63           &  $15.776$  &   $15.753$   &   $0.71$   &  $16.307$  &   $16.259$   &   $0.94$   &  $15.105$  &   $15.095$  &   $0.42$       \\
V64           &  $15.765$  &   $15.761$   &   $0.58$   &  $16.145$  &   $16.119$   &   $0.65$   &  $15.207$  &   $15.162$  &   $0.38$       \\
V65           &  $15.802$  &   $15.791$   &   $0.54$   &  $16.317$  &   $16.294$   &   $0.56$   &  $15.145$  &   $15.140$  &   $0.34$       \\
V66           &  $15.892$  &   $15.882$   &   $    $   &  $16.250$  &   $16.232$   &   $    $   &  $15.389$  &   $15.384$  &   $    $       \\
V67           &  $15.881$  &   $15.874$   &   $    $   &  $16.251$  &   $16.238$   &   $    $   &  $15.303$  &   $15.297$  &   $    $       \\
 & & & & & & & & & \\[-5pt]
V68           &  $15.773$  &   $15.757$   &   $0.50$   &  $16.185$  &   $16.165$   &   $0.65$   &  $15.181$  &   $15.167$  &   $    $       \\
V69           &  $16.002$  &   $15.977$   &   $1.03$   &  $16.536$  &   $16.499$   &   $1.16$   &  $15.236$  &   $15.224$  &   $0.58$       \\
V70           &  $15.854$  &   $15.844$   &   $0.49$   &  $16.279$  &   $16.265$   &   $0.53$   &  $15.217$  &   $15.209$  &   $0.38$       \\
V71           &  $15.884$  &   $15.870$   &   $0.60$   &  $16.301$  &   $16.274$   &   $0.73$   &  $15.238$  &   $15.225$  &   $    $       \\
V73           &  $15.767$  &   $15.753$   &   $    $   &  $16.176$  &   $16.169$   &   $0.43$   &  $15.216$  &   $15.199$  &   $    $       \\
 & & & & & & & & & \\[-5pt]
V74           &  $15.883$  &   $15.862$   &   $0.56$   &  $16.186$  &   $16.155$   &   $0.73$   &  $15.440$  &   $15.432$  &   $0.38$       \\
V75           &  $15.872$  &   $15.855$   &   $0.50$   &  $16.306$  &   $16.284$   &   $0.72$   &  $15.259$  &   $15.253$  &   $    $       \\
V76           &  $15.871$  &   $15.864$   &   $0.40$   &  $16.316$  &   $16.303$   &   $0.50$   &  $15.217$  &   $15.207$  &   $    $       \\
V77           &  $15.697$  &   $15.681$   &   $0.71$   &  $16.118$  &   $16.075$   &   $0.98$   &  $15.051$  &   $15.027$  &   $    $       \\
V78           &  $15.699$  &   $15.676$   &   $0.89$   &  $16.246$  &   $16.215$   &   $1.04$   &  $15.103$  &   $15.095$  &   $0.70$       \\
 & & & & & & & & & \\[-5pt]
V80           &  $15.720$  &   $15.697$   &   $0.72$   &  $16.265$  &   $16.228$   &   $0.86$   &  $15.058$  &   $15.045$  &   $    $       \\
V81           &  $15.812$  &   $15.791$   &   $    $   &  $16.348$  &   $16.325$   &   $    $   &  $15.061$  &   $15.055$  &   $    $       \\
V82           &  $15.664$  &   $15.656$   &   $    $   &  $16.154$  &   $16.144$   &   $0.55$   &  $15.043$  &   $15.027$  &   $    $       \\
V83           &  $15.587$  &   $15.578$   &   $0.95$   &  $16.223$  &   $16.187$   &   $1.04$   &  $15.002$  &   $14.985$  &   $    $       \\
V84           &  $16.031$  &   $15.989$   &   $1.05$   &  $16.491$  &   $16.430$   &   $1.12$   &  $15.435$  &   $15.415$  &   $    $       \\
& & & & & \\[-5pt]
V86           &  $13.669$  &   $13.659$   &   $    $   &  $14.380$  &   $14.368$   &   $    $   &  $12.659$  &   $12.646$  &   $    $       \\
V87           &  $15.961$  &   $15.943$   &   $0.65$   &  $16.264$  &   $16,236$   &   $0.75$   &            &             &                 \\
V88           &  $15.592$  &   $15.567$   &   $0.80$   &  $16.091$  &   $16.058$   &   $0.96$   &  $14.840$  &   $14.818$  &   $    $       \\
V89           &  $15.791$  &   $15.785$   &   $    $   &  $16.255$  &   $16.250$   &   $    $   &  $15.082$  &   $15.076$  &   $    $       \\
V90           &  $15.754$  &   $15.742$   &   $0.48$   &  $16.151$  &   $16.124$   &   $0.83$   &  $15.210$  &   $15.202$  &   $    $       \\
& & & & & \\[-5pt]
V92           &  $15.804$  &   $15.798$   &   $    $   &  $16.311$  &   $16.303$   &   $0.44$   &  $15.143$  &   $15.129$  &   $    $       \\
V93           &  $15.820$  &   $15.810$   &   $0.40$   &  $16.218$  &   $16.202$   &   $0.49$   &  $15.176$  &   $15.171$  &   $    $       \\
V97           &  $15.779$  &   $15.765$   &   $0.63$   &  $16.240$  &   $16.209$   &   $0.81$   &  $15.143$  &   $15.136$  &   $0.42$       \\
V111          &  $ $  &   $ $   &   $    $   &  $16.663$  &   $16.663$   &   $    $   &  $ $  &   $   $  &   $    $       \\
V117          &  $ $  &   $   $   &   $    $   &  $17.020$  &   $17.019$   &   $    $   &  $ $  &   $ $  &   $    $       \\
& & & & & \\[-5pt]
V118          &  $15.847$  &   $15.836$   &   $0.49$   &  $16.133$  &   $16.118$   &   $0.58$   &  $15.389$  &   $15.384$  &   $    $       \\
V119          &  $16.050$  &   $16.046$   &   $    $   &  $16.728$  &   $16.727$   &   $0.35$   &  $15.099$  &   $15.091$  &   $    $       \\
V130          &  $15.443$  &   $15.347$   &   $    $   &  $16.097$  &   $16.047$   &   $0.95$   &  $14.707$  &   $14.613$  &   $    $       \\
V131          &            &              &            &  $15.677$  &   $15.653$   &   $0.80$   &            &             &                \\
V139          &            &              &            &  $15.920$  &   $15.906$   &   $0.55$   &            &             &                \\
& & & & & \\[-5pt]
V141          &  $15.546$  &   $15.533$   &   $    $   &  $15.985$  &   $15.978$   &   $0.56$   &  $14.915$  &   $14.887$  &   $    $       \\
V160=ZK52     &  $15.808$  &   $15.741$   &   $    $   &  $15.903$  &   $15.894$   &   $0.60$   &  $15.133$  &   $14.863$  &   $    $       \\
V162=ZK18     &  $ $  &   $ $   &   $    $   &  $16.159$  &   $16.140$   &   $0.85$   &  $ $  &   $ $  &   $    $      \\
V163          &  $15.722$  &   $15.674$   &   $    $   &  $16.043$  &   $16.004$   &   $1.19$   &  $14.970$  &   $14.945$  &   $    $       \\
V166=ZK22     &  $15.447$  &   $15.424$   &   $0.53$   &  $15.884$  &   $15.866$   &   $0.67$   &  $14.767$  &   $14.754$  &   $    $       \\
& & & & & \\[-5pt]
V167          &  $15.538$  &   $15.531$   &   $    $   &  $15.940$  &   $15.931$   &   $0.60$   &  $14.764$  &   $14.759$  &   $    $       \\
V168          &  $15.734$  &   $15.717$   &   $    $   &  $16.230$  &   $16.201$   &   $0.93$   &  $15.005$  &   $14.994$  &   $    $       \\
V169=ZK23     &  $15.840$  &   $15.807$   &   $1.00$   &  $16.409$  &   $16.345$   &   $1.35$   &  $15.074$  &   $15.033$  &   $    $       \\
V170	      &  $15.735$  &   $15.713$   &   $0.63$   &  $16.248$  &   $16.225$   &   $0.77$   &  $15.141$  &   $15.112$  &   $    $       \\
V171          &  $15.818$  &   $15.801$   &   $    $   &  $16.294$  &   
$16.261$   &   $0.84$   &  $15.131$  &   $15.121$  &   $    $       \\
& & & & & \\[-5pt]
V172	      &  $15.113$  &   $15.112$   &   $0.20$   &  $15.964$  &   $15.958$   &   $0.35$   &  & \\
V173=ZK11     &  $15.748$  &   $15.737$   &   $0.46$   &  $16.212$  &   $16.192$   &   $0.69$   &  $15.053$  &   $15.048$  &   $    $       \\
V177=ZK64     &  $15.585$  &   $15.578$   &   $0.32$   &  $16.075$  &   $16.066$   &   $0.54$   &  $14.900$  &   $14.891$  &   $    $       \\
V178=ZK55     &  $15.750$  &   $15.741$   &   $0.34$   &  $16.212$  &   $16.199$   &   $0.48$   &  $15.142$  &   $15.124$  &   $    $       \\
V181=ZK14     &  $ $  &   $ $   &   $    $   &  $16.015$  &   $16.013$   &   $    $   &  $ $  &   $   $  &   $    $       \\
& & & & & \\[-5pt]
ZK4           &  $15.812$  &   $15.801$   &   $0.45$   &  $16.222$  &   $16.209$   &   $0.53$   &  $15.161$  &   $15.151$  &   $    $       \\
ZK5           &  $15.767$  &   $15.751$   &   $0.55$   &  $16.181$  &   $18.173$   &   $0.63$   &  $15.284$  &   $15.273$  &   $    $       \\
ZK34          &  $15.819$  &   $15.760$   &   $    $   &  $16.218$  &   $16.202$   &   $0.70$   &  $15.015$  &   $14.929$  &   $    $       \\
ZK63          &  $15.594$  &   $15.587$   &   $0.35$   &  $16.014$  &   $16.005$   &   $0.50$   &  $15.007$  &   $15.003$  &   $    $       \\
ZK67          &  $ $  &   $ $   &   $    $   &  $16.463$  &   $16.340$   &   $1.40$   &  $ $  &   $ $  &   $    $       \\
& & & & & \\[-5pt]
ZK69          &  $ $  &   $ $   &   $    $   &  $16.090$  &   $16.085$   &   $0.40$   &  $ $  &   $ $  &   $    $       \\
ZK74          &  $15.805$  &   $15.787$   &   $0.53$   &  $16.135$  &   $16.109$   &   $0.88$   &  $15.293$  &   $15.284$  &   $    $       \\
ZK78          &  $15.737$  &   $15.727$   &   $0.36$   &  $16.120$  &   $16.117$   &   $0.54$   &  $15.159$  &   $15.156$  &   $0.23$       \\
ZK80          &  $15.789$  &   $15.781$   &   $0.37$   &  $16.237$  &   $16.223$   &   $0.54$   &  $15.163$  &   $15.159$  &   $    $       \\
NV1           &  $17.950$  &   $17.933$   &   $    $   &  $18.338$  &   $18.325$   &   $0.60$   &  $17.412$  &   $17.400$  &   $    $       \\
& & & & & \\[-5pt]
NV2           &  $ $  &   $ $   &   $    $   &  $16.236$  &   $16.229$   &   $    $   &  $ $  &   $ $  &   $    $       \\
NV3           &  $15.671$  &   $15.658$   &   $0.40$   &  $16.083$  &   $16.072$   &   $0.55$   &  $15.151$  &   $15.137$  &   $    $       \\
NV5           &  $15.760$  &   $15.725$   &   $0.85$   &  $16.241$  &   $16.225$   &   $1.13$   &  $15.107$  &   $15.012$  &   $    $       \\
NV6           &  $15.779$  &   $15.773$   &   $0.32$   &  $16.192$  &   $16.181$   &   $0.42$   &  $15.220$  &   $15.217$  &   $0.25$       \\
NV7           &  $15.740$  &   $15.731$   &   $0.38$   &  $16.142$  &   $16.125$   &   $0.52$   &  $15.148$  &   $15.143$  &   $0.26$       \\
& & & & & \\[-5pt]
NV9           &  $15.826$  &   $15.813$   &   $0.45$   &  $16.370$  &   $16.349$   &   $0.56$   &  $15.158$  &   $15.147$  &   $    $       \\
NV10          &  $15.792$  &   $15.782$   &   $0.35$   &  $16.292$  &   $16.281$   &   $0.45$   &  $15.151$  &   $15.147$  &   $0.24$       \\
NV11          &  $ $  &   $ $   &   $    $   &  $16.266$  &   $16.257$   &   $    $   &  &   $ $  &   $    $       \\
NV12          &  $ $  &   $ $   &   $    $   &  $16.169$  &   $16.155$   &   $    $   &  $ $  &   $ $  &   $    $       \\
NV13          &  $ $  &   $ $   &   $    $   &  $16.077$  &   $16.075$   &   $0.56$   &  $ $  &   $ $  &   $    $       \\
& & & & & \\[-5pt]
NV14          &  $15.812$  &   $15.807$   &   $0.31$   &  $16.090$  &   $16.081$   &   $0.43$   &  $15.339$  &   $15.333$  &   $    $       \\
\enddata
\end{deluxetable}

\end{document}